\documentclass[%
 aip,
 amsmath,amssymb,
 reprint,%
]{revtex4-2}

\usepackage{graphicx}
\graphicspath{{Figures/}}
\usepackage{dcolumn}
\usepackage{bm}
\usepackage[colorlinks, allcolors=blue]{hyperref}

\usepackage{titlesec}
\titleformat*{\section}{\large\bfseries}

\usepackage[utf8]{inputenc}
\usepackage[T1]{fontenc}
\usepackage{mathptmx}
\usepackage{amsmath}

\usepackage{siunitx}
\usepackage{blindtext}

\usepackage[acronym]{glossaries}
\glsdisablehyper
\setacronymstyle{long-short}

\newacronym{mocvd}{MOCVD}{metal-organic chemical vapour deposition}
\newacronym{movpe}{MOVPE}{metalorganic vapour-phase epitaxy}
\newacronym{cl}{CL}{cathodoluminescence}
\newacronym{trcl}{TRCL}{time-resolved cathodoluminescence}
\newacronym{pl}{PL}{photoluminescence}
\newacronym{trpl}{TRPL}{time-resolved PL}
\newacronym{afm}{AFM}{atomic force microscopy}
\newacronym{kpfm}{KPFM}{Kelvin probe force microscopy}
\newacronym{tem}{TEM}{transmission electron microscopy}
\newacronym{tipl}{TIPL}{time-integrated PL}
\newacronym{stem}{STEM}{scanning transmission electron microscopy}
\newacronym{eds}{EDX}{energy-dispersive X-ray spectroscopy}
\newacronym{haadf}{HAADF}{high-angle annular dark-field}
\newacronym{log}{LoG}{Laplacian of Gaussian}
\newacronym{dlos}{DLOS}{deep level optical spectroscopy}

\newacronym{led}{LED}{light-emitting diode}
\newacronym{ld}{LD}{laser diode}
\newacronym[longplural = internal quantum efficiencies]{iqe}{IQE}{internal quantum efficiency}
\newacronym{ul}{UL}{"underlayer"}
\newacronym{sl}{SL}{superlattice}
\newacronym{pd}{PD}{point defect}
\newacronym{upd}{UPD}{UL-interacting PD}
\newacronym{sd}{SD}{surface defect}
\newacronym{qw}{QW}{quantum well}
\newacronym{sqw}{SQW}{single quantum well}
\newacronym{fs}{FS}{freestanding}
\newacronym{cw}{CW}{continuous-wave}
\newacronym{ccd}{CCD}{charge-coupled device}
\newacronym{irf}{IRF}{instrument response function}
\newacronym{sem}{SEM}{scanning electron microscopy}
\newacronym{se}{SE}{secondary electron}
\newacronym{ml}{ML}{monolayer}
\newacronym{s}{S}{single}
\newacronym{cv}{CV}{capacitance-voltage}
\newacronym{srh}{SRH}{Shockley-Read-Hall}
\newacronym{nrc}{NRC}{non-radiative recombination centre}
\newacronym{lo}{LO}{longitudinal-optical}
\newacronym[longplural = full width at half maxima]{fwhm}{FWHM}{full width at half maximum}

\begin{document}

\title[]{Imaging non-radiative point defects buried in quantum wells using cathodoluminescence}

\author{T. F. K. Weatherley}
\email{thomas.weatherley@epfl.ch}	
\affiliation{ \mbox{Institute of Physics, Ecole Polytechnique Fédérale de Lausanne (EPFL), CH-1015 Lausanne, Switzerland}}%

\author{W. Liu}
\affiliation{ \mbox{Institute of Physics, Ecole Polytechnique Fédérale de Lausanne (EPFL), CH-1015 Lausanne, Switzerland}}%
\author{V. Osokin}
\affiliation{\mbox{Department of Physics, The Clarendon Laboratory, University of Oxford, Oxford OX1 3PU, UK}}%
\author{D. T. L. Alexander}
\affiliation{ \mbox{Institute of Physics, Ecole Polytechnique Fédérale de Lausanne (EPFL), CH-1015 Lausanne, Switzerland}}%

\author{R. A. Taylor}
\affiliation{\mbox{Department of Physics, The Clarendon Laboratory, University of Oxford, Oxford OX1 3PU, UK}}%
\author{J.-F. Carlin}
\affiliation{ \mbox{Institute of Physics, Ecole Polytechnique Fédérale de Lausanne (EPFL), CH-1015 Lausanne, Switzerland}}%
\author{R. Butté}
\affiliation{ \mbox{Institute of Physics, Ecole Polytechnique Fédérale de Lausanne (EPFL), CH-1015 Lausanne, Switzerland}}%
\author{N. Grandjean}
\affiliation{ \mbox{Institute of Physics, Ecole Polytechnique Fédérale de Lausanne (EPFL), CH-1015 Lausanne, Switzerland}}%

\date{\today}

\pdfbookmark[1]{Abstract}{sec:abs}

\begin{abstract}	
	Crystallographic point defects (PDs) can dramatically decrease the efficiency of optoelectronic semiconductor devices, many of which are based on quantum well (QW) heterostructures. However, spatially resolving \textit{individual} non-radiative PDs buried in such QWs has so far not been demonstrated. Here, using high-resolution cathodoluminescence (CL) and a specific sample design, we spatially resolve, image, and analyse non-radiative PDs in InGaN/GaN QWs. We identify two different types of PD by their contrasting behaviour with temperature, and measure their densities from $10^{14}$ \si{\per\centi\metre\cubed} to as high as $10^{16}$ \si{\per\centi\metre\cubed}. Our CL images clearly illustrate the interplay between PDs and carrier dynamics in the well: increasing PD concentration severely limits carrier diffusion lengths, while a higher carrier density suppresses the non-radiative behaviour of PDs. The results in this study are readily interpreted directly from CL images, and represent a significant advancement in nanoscale PD analysis.
\end{abstract}

\maketitle

\pdfbookmark[1]{Introduction}{sec:intro}

\noindent Crystallographic \glspl{pd} have a profound effect on the optical properties of semiconductors.\cite{Bourgoin1983} Of particular interest are \glspl{pd} with energy levels deep in the bandgap, which can act as efficient electron-hole recombination centres.\cite{Shockley1952} Recombination at \glspl{pd} can be useful: nitrogen impurities increase the efficiency of gallium-phosphide based \glspl{led},\cite{Thomas1965, Wight1977} while silicon/nitrogen-vacancy centres in diamond are promising single photon emitters.\cite{Aharonovich2016, Bradac2019} However, it is more common for \gls{pd} recombination to result in no light emission at all, and such non-radiative \glspl{pd} can dramatically decrease the \gls{iqe} of optoelectronic devices such as \glspl{led}, laser diodes, and solar cells.\cite{Kressel1981, Ball2016, Park2018} Despite the relevance of these defects to both industry and research, spatially resolving, imaging, and analysing \textit{individual} \glspl{pd} remains a serious challenge, particularly at high defect densities. Previous studies have analysed surface defects or \glspl{pd} in 2D materials,\cite{Bourrellier2016, Sang2016, Jiang2018, Feng2018, Edelberg2019, Ziatdinov2019} but have not imaged \glspl{pd} buried in semiconductor heterostructures due to the difficulty of pinpointing atomic-scale defects in bulk. As a consequence, \glspl{pd} have not been individually imaged in \glspl{qw}, even though such wells are the active region in the majority of today's \glspl{led} and laser diodes. Imaging the non-radiative \glspl{pd} that can plague these devices is particularly difficult since they lack any localised luminescence, ruling out the use of super-resolution techniques.\cite{Rittweger2009,Feng2018}

The critical role of non-radiative \glspl{pd} in \glspl{qw} is exemplified perfectly by III-nitride semiconductors: recent literature indicates that an intrinsic \gls{pd} in InGaN/GaN \glspl{qw} acts as a highly-effective non-radiative recombination centre, killing the \gls{iqe} of green to near-ultraviolet \glspl{led}.\cite{Armstrong2012, Armstrong2014, Armstrong2015, Haller2017, Haller2018} These \glspl{pd} arise during growth from an initial population of GaN surface defects which are only incorporated into layers with indium content.\cite{Haller2018} Growing an indium-containing "underlayer" below the \gls{qw} can therefore trap the surface defects before they incorporate into the well. Varying the thickness of this underlayer allows for precise \gls{pd} density control in the \gls{qw},  making III-nitrides the ideal platform to explore non-radiative \gls{pd} imaging. III-nitrides are also the focus of ongoing research in deep-ultraviolet \glspl{led},\cite{Shatalov2017, Kneissl2019} green laser diodes,\cite{Ohta2010,Adachi2014} and micro-\glspl{led}.\cite{Lin2020}

Here, using \gls{cl}, we spatially resolve, image, and analyse individual non-radiative \glspl{pd} in InGaN/GaN \glspl{qw} up to densities as high as $10^{16}$ \si{\per\centi\metre\cubed}. We identify two different types of \glspl{pd} in the \glspl{qw}, and show that one type, linked to a mid-gap deep state, decreases the room temperature \gls{iqe} from around 64 \% to only 1 \% with increasing density. Beyond just mapping non-radiative \glspl{pd}, we also use \gls{cl} images to explore their interaction with carrier dynamics in the well. We evidence the direct impact of \glspl{pd} on carrier diffusion lengths, and demonstrate how higher carrier densities strongly reduce the influence of \glspl{pd} on the \gls{cl} intensity.


\begin{figure*}[t]
	\centering
	\includegraphics[width=1\textwidth]{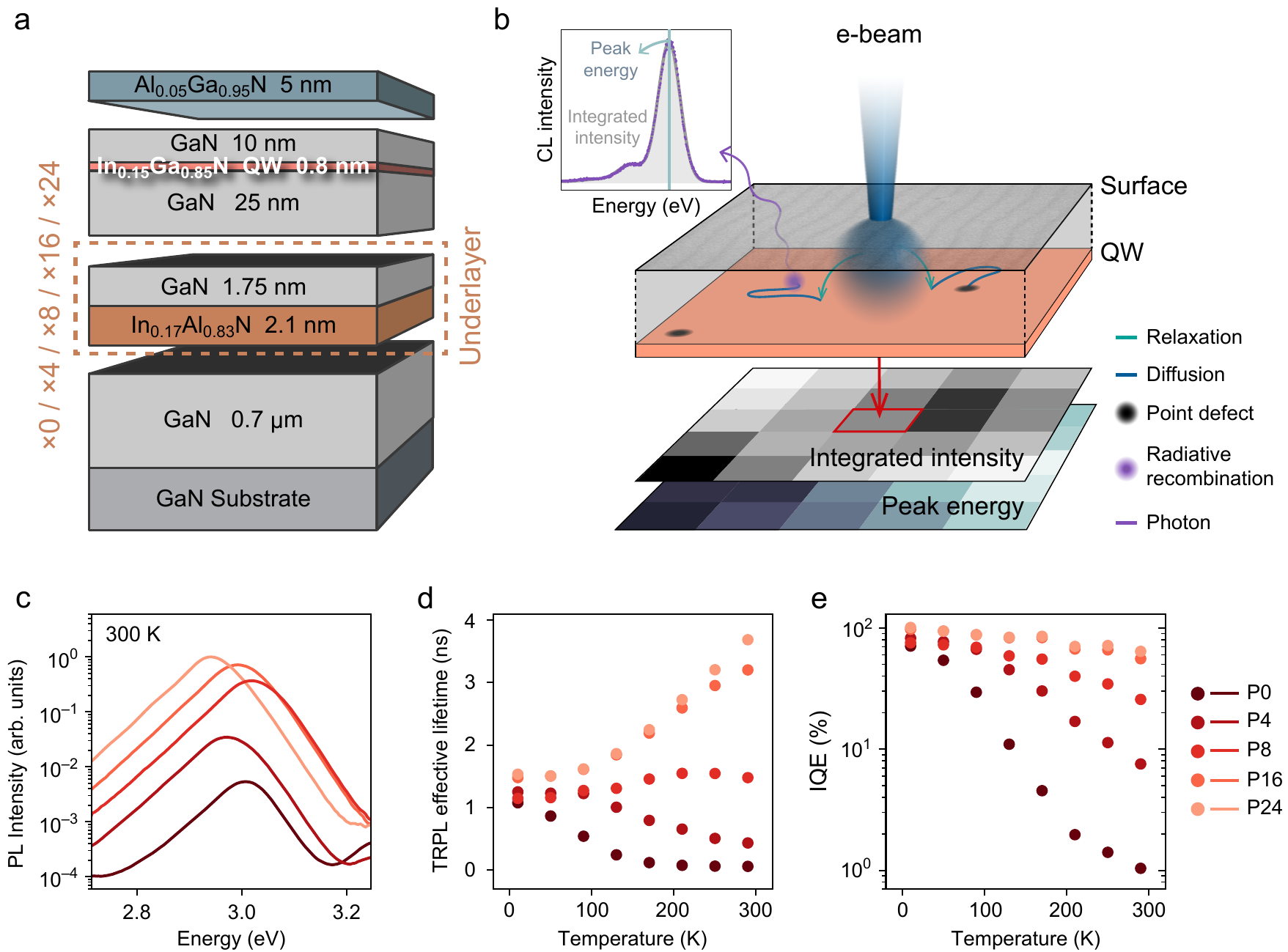}
	\caption{{\fontfamily{phv}\selectfont\footnotesize Schematics and initial characterisation of the samples. \textbf{a} Schematic of the sample design---five samples were prepared with 0, 4, 8, 16, and 24 periods in the SL; these samples are referred to as P0--P24, respectively. A thin \glsentryshort{qw} allows for shorter diffusion lengths, and the top GaN layer is only 10 nm to excite close to the \glsentryshort{qw}. \textbf{b} Diagram depicting \glsentryshort{cl} imaging of a \glsentryshort{qw}. Examples of possible carrier paths from one excitation position are illustrated, with transfer to the \glsentryshort{qw} followed by radiative recombination or recombination at a \glsentryshort{pd}. Photons are globally collected to generate a \gls{cl} spectrum for this position (example top left). This process is repeated as the beam is scanned over the area of interest, and all collected spectra are fitted to generate \glsentryshort{qw} integrated intensity and peak energy images of this area. \glsentryshortpl{pd} present as dark areas in the intensity image since they locally reduce radiative recombination. \textbf{c} Room temperature \glsentryshort{pl} spectra showing the \glsentryshort{qw} emission of all samples; peak intensities scale with the underlayer thickness. \textbf{d} The relation of effective carrier lifetime to temperature for each sample gained from \glsentryshort{trpl}, along with \textbf{e} corresponding temperature-dependent \glsentryshort{iqe} extracted from these effective lifetimes. Effective lifetimes and \glsentryshortpl{iqe} increase for samples with thicker underlayers, indicating a reduction in defect density.}}
	\label{sample_info}
\end{figure*}

\pdfbookmark[1]{Sample preparation and macroscale characterisation}{sec:sample}

\vspace{0.5cm}
\noindent\fontfamily{phv}\textbf{Results}\\\normalfont
\textbf{Sample preparation and macroscale characterisation.} We grew our samples by metal-organic vapour phase epitaxy, using \textit{c}-plane freestanding GaN substrates (dislocation density  $\sim 10^6$ \si{\per\centi\metre\squared}) to minimise the influence of threading dislocations on our results. A lattice-matched In\textsubscript{0.17}Al\textsubscript{0.83}N/GaN \gls{sl} was chosen as an underlayer to bury surface defects.\cite{Haller2018, Haller2019a} Five samples with varying underlayer thickness were grown by changing the number of \gls{sl} periods, $n$---these samples are then labelled as P$n$ (Fig.\ \ref{sample_info}a). An Al\textsubscript{0.05}Ga\textsubscript{0.95}N cap is included to avoid surface recombination. 

These samples are tailored for high-resolution \gls{cl}. During \gls{cl} acquisition, the electron beam is raster-scanned over the sample surface, pausing regularly to gather data at evenly-spaced positions. At each position, the sample is locally excited while luminescence is globally collected---the full process is depicted in Fig.\ \ref{sample_info}b. As the electron beam strikes the surface, electron-hole pairs generated within the excitation volume rapidly relax to the \gls{qw} (green arrows in Fig.\ \ref{sample_info}b), diffusing laterally (blue paths) until they reach a \gls{pd}, or until they undergo radiative recombination. With continuous excitation, the carrier density within the \gls{qw} reaches a steady-state spatial distribution, and photons (purple arrow) are emitted from radiative recombinations within this distribution. These photons are collected and dispersed in a spectrometer to obtain a \gls{cl} spectrum for each excitation position in the scan---so called "hyperspectral" imaging (an example spectrum is shown in Fig.\ \ref{sample_info}b). By fitting the \gls{qw} emission at every position in this full spectral dataset (see supplementary Sec.\ \ref{hyp_fit_sec}), we generate both a \gls{qw} integrated intensity image and a peak emission energy image of the scanned area.

Since photons are globally collected, the \gls{cl} spectrum recorded for one position is spatially averaged over a \gls{qw} area related to (i) the carrier diffusion length and (ii) the excitation volume. These parameters are often not controlled in \gls{qw} \gls{cl} studies, which can limit the spatial resolution to hundreds of nanometres or more. Here, we discuss how we minimise both parameters in our samples to obtain high-resolution images.

\begin{figure*}[t]
	\centering
	\includegraphics[width=1\textwidth]{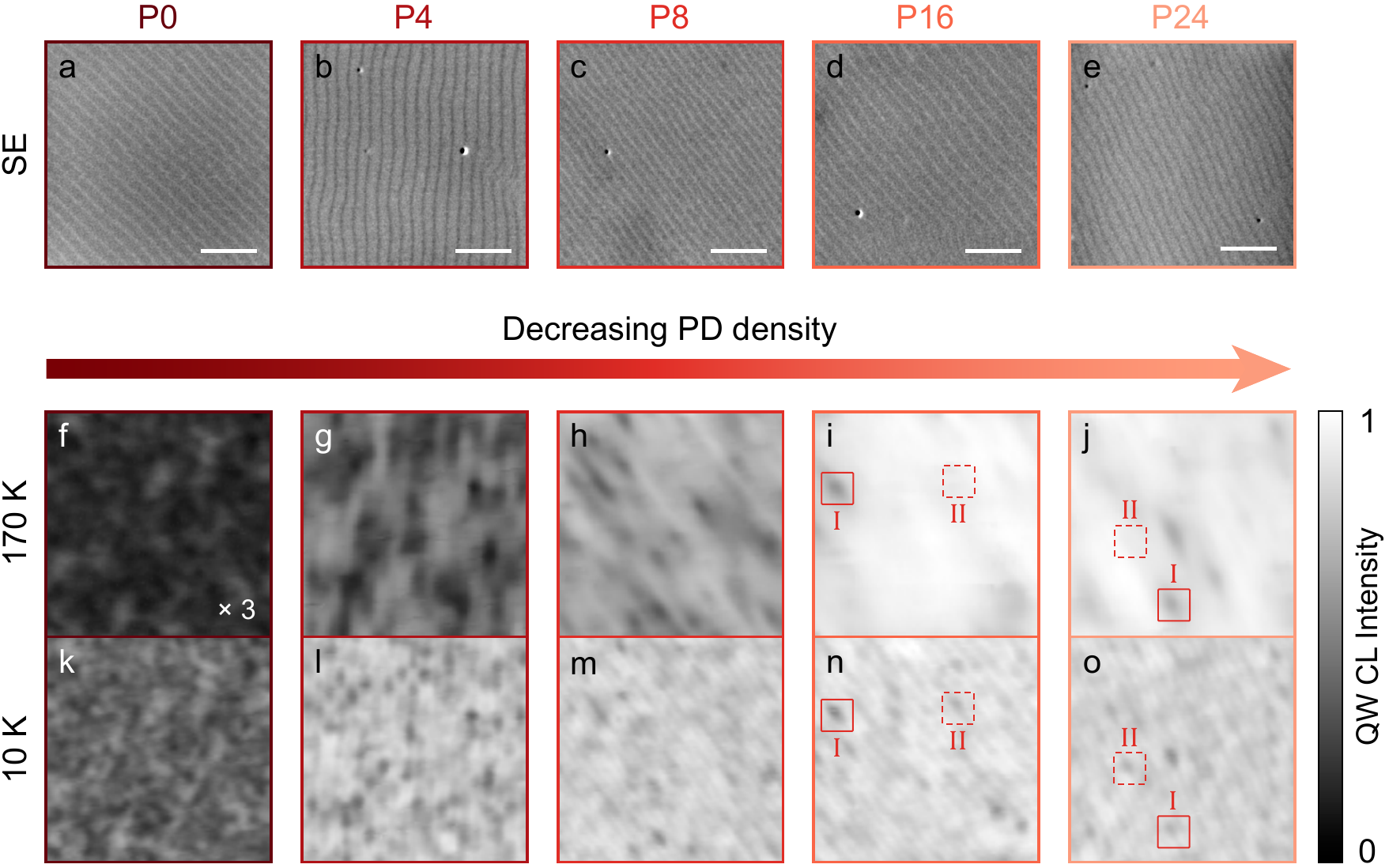}
	\caption{{\fontfamily{phv}\selectfont\footnotesize \glsentryshort{se} images and temperature-dependent \glsentryshort{cl} for all samples. \textbf{a -- e} \glsentryshort{se} images for each sample (P0 -- P24); step-edges are clear as parallel lines and V-shaped pits marking the position of threading dislocations are visible as dark spots. All are $2 \times 2$ \si{\micro\metre\squared} with a 500 nm scale bar. Meanwhile, \textbf{f -- j} and \textbf{k -- o} show the \glsentryshort{qw} \glsentryshort{cl} integrated intensity images acquired \textit{on the same $2 \times 2$ \si{\micro\metre\squared} area} for each sample at 170 K and 10 K, respectively. Dark regions are a direct consequence of non-radiative recombination at \glsentryshortpl{pd}, and decrease in density as the \glsentryshort{sl} periods are increased. Each image is plotted on an absolute intensity scale from 0 -- 1 (normalised to the highest intensity point in P24). The red squares in \textbf{i}, \textbf{j}, \textbf{n}, and \textbf{o} indicate examples of type-I \glsentryshortpl{pd} with a strong impact on high-temperature intensity (solid line), and type-II \glsentryshortpl{pd} with a lesser impact (dashed line). All images obtained at a probe current of 200 pA. The 170 K intensity for P0 has been multiplied by three to make the image visible.}}
	\label{pd_cl}
	
\end{figure*} 

Carrier diffusion lengths can be reduced by lower diffusion coefficients and shorter carrier lifetimes. The diffusion coefficient in \mbox{InGaN/GaN} \glspl{qw} is mainly controlled by temperature, starting from near-zero at 10 K and increasing until it saturates at around 200 K.\cite{Solowan2013} Consequently, we reach the highest resolution at cryogenic temperatures. We also decrease diffusion lengths by choosing a very thin single \gls{qw} of only three molecular monolayers ($\sim$ 0.8 nm, Fig. \ref{sample_info}a), which reduces the carrier radiative lifetime in the well.\cite{Berkowicz2000} 

To minimise the excitation volume we use an electron-beam acceleration voltage of only 1.5 kV. At this voltage 95 \% of the incident energy is absorbed within the top 20 nm of the sample (supplementary Sec.\ \ref{carr_inj_sec}), so we select a thin 10 nm top GaN layer to excite as close to the \gls{qw} as possible (Fig.\ \ref{sample_info}a). 

Before moving on to \gls{cl} results, let us initially gauge the macroscale properties of our samples by more conventional methods for comparison. The room-temperature \gls{pl} spectra shown in Fig.\ \ref{sample_info}c clearly demonstrate the rise in \gls{qw} efficiency achieved by the underlayer, with the peak intensity increasing by over two orders of magnitude from P0 to P24 (minor shifts in peak energy between different samples are explained in supplementary Sec.\ \ref{tem_sec}). This efficiency improvement is underscored by the striking contrast in \gls{qw} effective lifetimes, gained from fitting \gls{trpl} decay curves at early delays (Fig.\ \ref{sample_info}d) (see supplementary Sec.\ \ref{trpl_sec} for decay curves). The lifetime behaviour of P24 indicates it is dominated by radiative recombination across the full temperature range, since it exactly matches radiative lifetime behaviour in an ideal \gls{qw}.\cite{Matsusue1987, Feldmann1987, Andreani1991, Hangleiter2017} For other samples, as the number of \gls{sl} periods is reduced the high-temperature effective lifetimes decrease due to thermal activation of defect-assisted recombination.\cite{Henry1977, Hangleiter2018} The difference in \gls{iqe} between the samples can be quantified with a simple treatment comparing the initial \gls{trpl} decay curve intensity to the effective lifetime (supplementary Sec.\ \ref{trpl_sec}).\cite{Langer2013} Results are displayed in Fig.\ \ref{sample_info}e: as expected, P24 maintains a high \gls{iqe} of 64 $\pm$ 3 \%  at room temperature, which matches that measured for a similar sample in another study.\cite{Haller2019a} Samples with fewer \gls{sl} periods present much lower efficiencies, with the \gls{iqe} of P0 falling to only $\sim1$ \% at room temperature.  Overall, the results presented in Figs.\ \ref{sample_info}c-e confirm that the defect density in the \mbox{InGaN/GaN} single \gls{qw} significantly reduces on going from P0 to P24. With only these measurements, however, it is very difficult to precisely quantify the defect density, let alone identify their nature.

\vspace{0.5cm}

\pdfbookmark[1]{Cathodoluminescence}{sec:CL}

\noindent\textbf{Cathodoluminescence.} To spatially resolve and directly analyse the \glspl{pd}, we turn our attention to \gls{cl} results (Fig.\ \ref{pd_cl}). Atomic-height step-edges are noticeable in the $2 \times 2$ \si{\micro\metre\squared} \gls{se} images (Figs.\ \ref{pd_cl}a--e) as a series of parallel lines; all samples present the same terraced morphology, indicating ideal step-flow growth (see supplementary Sec.\ \ref{afm_sec} for complementary atomic force microscopy). In addition, very few dislocation-induced V-shaped pits are visible due to the growth on freestanding GaN substrates.

We focus on \gls{cl} integrated intensity images obtained by fitting the \gls{qw} emission from hyperspectral data (Figs.\ \ref{pd_cl}f--o). Two temperatures were chosen for \gls{cl} characterisation: 10 K, at which the diffusivity in the well is near zero so resolution is only limited by the excitation volume,\cite{Solowan2013} and 170 K, which allows us to probe the high-temperature regime while retaining sufficient signal-to-noise ratio for the most defective samples. The images obtained at 10 K and 170 K for each sample are from precisely \textit{the same area} displayed in the \gls{se} images. The electron-beam probe current was set to 200 pA, which results in a steady-state carrier density near the peak-\gls{iqe} condition for high quality InGaN/GaN \glspl{qw} (see \hyperref[carr_sec]{carrier density section} and supplementary Sec.\ \ref{carr_inj_sec}).

We observe the impact from defects as local regions of low intensity in the 170 K results (Figs.\ \ref{pd_cl}f--j). Since dislocations identified in the \gls{se} images do not correlate with the dark regions, we attribute this low intensity to non-radiative recombination at \glspl{pd} buried in the \gls{qw}. The density of dark spots clearly decreases with increasing \gls{sl} periods, directly indicating a decline in \gls{pd} density in accordance with the \gls{pd}-reducing behaviour of the underlayer.\cite{Haller2018} Thus these \gls{cl} images reveal the nanoscale origin of the increasing \gls{iqe} from P0 to P24 in Fig.\ \ref{sample_info}e. The elongated shape of \gls{pd} dark regions in the less defective samples (Figs.\ \ref{pd_cl}h--j) is a result of asymmetric diffusion in the \glspl{qw} as discussed further in the \hyperref[diff_sec]{diffusion analysis section} of this paper.

Upon cooling to 10 K (Figs.\ \ref{pd_cl}k--o), the impact from \glspl{pd} is noticeably suppressed, as expected since the reduction in diffusion length means less carriers can reach \glspl{pd}. Non-radiative recombination at defects also usually requires multiple-phonon emission, which is less likely at low temperatures.\cite{Henry1977} Nevertheless, we still see some influence of \glspl{pd} on the \gls{qw} intensity. Observing effects from defects at cryogenic temperatures is not surprising: even at 10 K the \glspl{iqe} of the more defective samples are only $\sim$ 70 \% rather than 100 \% (Fig.\ \ref{sample_info}e). Such non-radiative recombination at low temperatures has previously been explained through tunneling-assisted transitions to defect levels.\cite{Hangleiter2018} With the improved spatial resolution of under 90 \si{\nano\metre} at this temperature (supplementary Sec.\ \ref{carr_inj_sec}), we can be confident we are observing individual \glspl{pd} as long as densities remain below $10^{17}$ \si{\per\centi\meter\cubed}.

Comparing the 10 K and 170 K results for P16 and P24 (Figs.\ \ref{pd_cl}i, j, n, \& o), we can distinguish two different types of \gls{pd}: one type leads to greater local intensity reduction at high temperature (type-I, solid red box), while the other has much less effect (type-II, dashed red box). With a clear decrease in the number of dark regions at 170 K with increasing \gls{sl} periods, we can already suggest the type-I defect density is reduced by the underlayer. However, actually quantifying the density of each type of defect requires a more detailed analysis of the \gls{cl} images.

\begin{figure}[t]
	\centering
	\includegraphics[width=\linewidth]{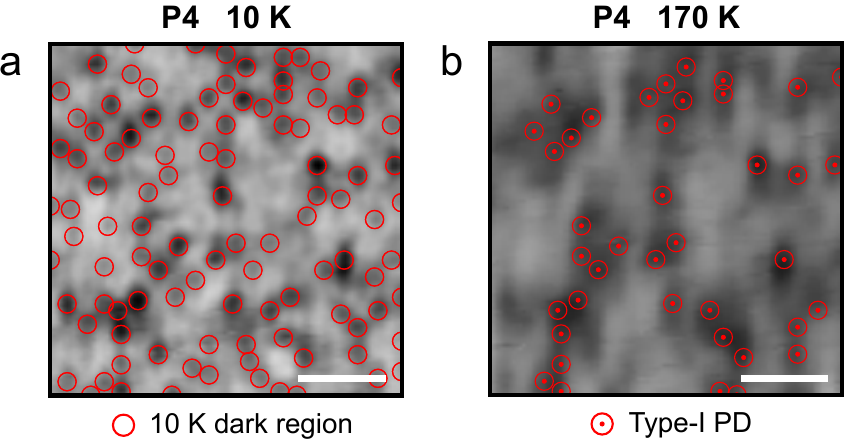}
	\caption{{\fontfamily{phv}\selectfont\footnotesize \glsentryshort{pd} detection example. \textbf{a} P4 10 K \glsentryshort{cl} intensity image from Fig.\ \ref{pd_cl} with red circles enclosing dark regions detected by the \glsentryshort{log} method (intensity scale 0.4--1). The detected dark regions which lead to an intensity at 170 K below a set limit, $I_\textnormal{lim}$, are identified as type-I \glspl{pd} as shown in \textbf{b} (intensity scale 0--1, $I_\textnormal{lim} = 0.32$). Scale bars are 500 nm.}}
	\label{pd_det}
	
\end{figure}

\begin{figure}[t]
	\centering
	\includegraphics[width=1\linewidth]{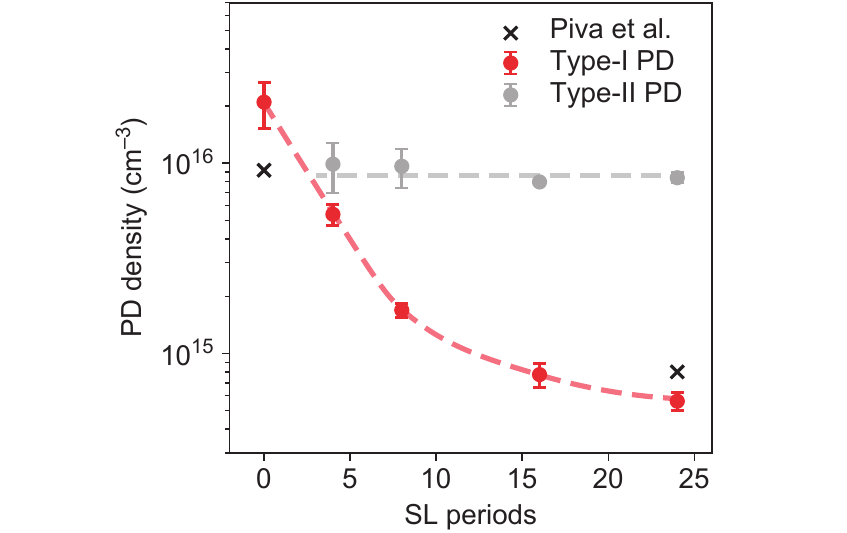}
	\caption{{\fontfamily{phv}\selectfont\footnotesize Type-I and type-II \glsentryshort{pd} densities in the \glsentryshort{qw} for each sample, demonstrating the drastic decrease in type-I \glsentryshort{pd} density with increasing underlayer thickness. Densities of mid-gap \glsentryshortpl{pd} calculated by Piva \textit{et al.} \protect\cite{Piva2020} on very similar samples to P0 and P24 are also displayed. Dashed lines are guides to the eye.}}
	\label{pd_dens}
\end{figure}

\begin{figure*}[t]
	\centering
	\includegraphics[width=1\linewidth]{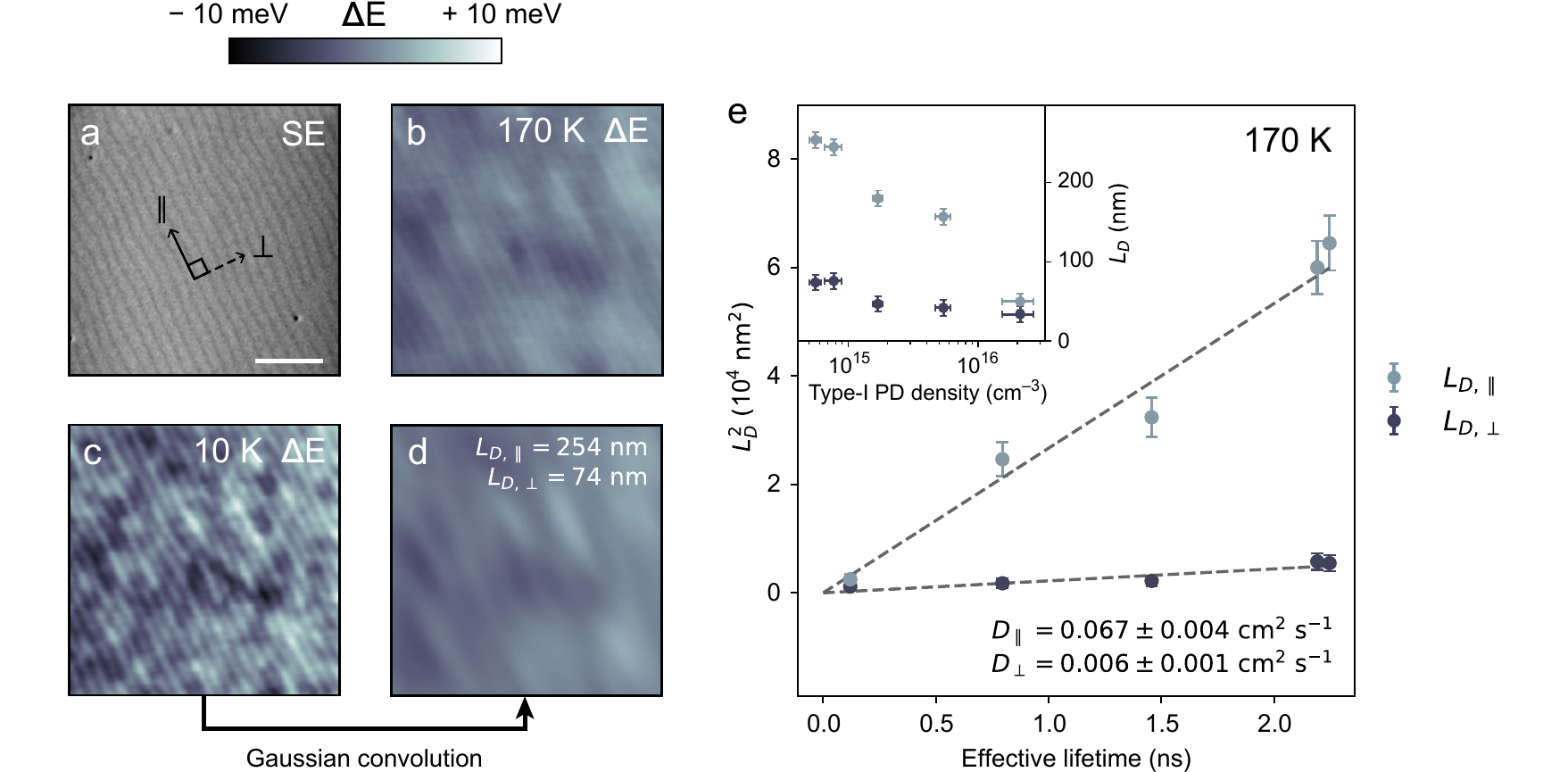}
	\caption{{\fontfamily{phv}\selectfont\footnotesize Diffusion analysis using peak energy of \glsentryshort{qw} emission. \textbf{a} \glsentryshort{se} image for P24 as in Fig.\ \ref{pd_cl}e, with directions parallel ($\parallel$) and perpendicular ($\perp$) to the step-edges highlighted (scale bar 500 nm). \textbf{b} \& \textbf{c} Images of \glsentryshort{qw} peak energy from the same area as in \textbf{a} at 170 K and 10 K, respectively. Energy modulation periodic with the step-edges is clear, and this modulation restricts carrier diffusion perpendicular to the step-edges. Restricted diffusion is confirmed by \textbf{d} the image calculated by Gaussian convolution fitting of the 10 K data in \textbf{c} to the 170 K data in \textbf{b}, since the extracted perpendicular diffusion length ($L_{D, \perp}$) is much shorter than the parallel length ($L_{D, \parallel}$). \textbf{e} Both diffusion lengths squared plotted against the effective lifetime of each sample. Linear fits are used to calculate the diffusivities at 170 K, $D_\parallel$ and $D_\perp$; the large difference in these diffusion coefficients emphasises the anisotropic diffusion in our \glsentryshortpl{qw}. The reduced chi-squared for these fits were 1.5 and 0.9 for parallel and perpendicular diffusion length data, respectively. The inset shows the severe shortening of diffusion lengths at 170 K with increasing type-I \gls{pd} density.}}
	\label{diff}
\end{figure*}

\vspace{0.5cm}

\pdfbookmark[1]{PD identification and counting}{sec:PD}

\noindent\textbf{\gls{pd} identification and counting.}  To estimate \gls{pd} densities, we need to differentiate the two kinds of \gls{pd} using their contrasting high-temperature behaviour as a criterion. This task is made more challenging by the loss of spatial resolution at 170 K due to increased carrier diffusion. Nonetheless, we can use 10 K images to accurately determine the position of \glspl{pd} since diffusion length is near zero at this temperature. Therefore, we established a general two-step procedure: (i) use the \gls{log} method\cite{Xiong2013} to automatically detect dark regions in a 10 K image, then (ii) superimpose the positions of these regions on a 170 K CL image of the same sample area. If a position is associated with an intensity below a defined limit at 170 K, $I_\textnormal{lim}$, it is identified as a type-I defect due to their strong impact at high temperature. For more information on our choice of reasonable $I_\textnormal{lim}$ and adapting the method for low defect densities in P16/P24, see supplementary Sec.\ \ref{pd_sec}. An example of the procedure applied to P4 is shown in Fig.\ \ref{pd_det}. While this procedure may not detect and correctly identify every \gls{pd}, its design ensures reproducibility and avoids bias arising from counting/differentiating \glspl{pd} manually. Once we have the 2-D defect densities from this method, we can calculate the 3-D densities using the full-width-at-half-maximum of the \gls{qw} indium profile measured by scanning transmission electron microscopy (supplementary Sec.\ \ref{tem_sec}), since the \glspl{pd} are contained within the well.

The final calculated \gls{pd} densities are shown in Fig.\ \ref{pd_dens}. We note that for P0 the very high concentration of type-I \glspl{pd} overwhelmed any effect from type-II \glspl{pd}, preventing the estimation of the type-II defect density. With a decrease in the concentration of type-I defects in the \gls{qw} from $10^{16}$ \si{\per\centi\metre\cubed} to only $10^{14}$ \si{\per\centi\metre\cubed} with increasing \gls{sl} periods, we propose that type-I \glspl{pd} are the key defects incorporated into the underlayer. In contrast, the concentration of type-II \glspl{pd} is rather insensitive to the underlayer thickness. We conclude that the drastic increase in room-temperature \gls{iqe} from P0 to P24 (Fig.\ \ref{sample_info}(e)) is entirely due to the reduction in type-I \gls{pd} density, while type-II defects have only a minor effect. The strong impact of type-I \glspl{pd} on \gls{cl} intensity and \gls{iqe} indicates they possess energy levels near the middle of the InGaN bandgap, since such midgap states are known to be efficient non-radiative recombination centres.\cite{Shockley1952, Dreyer2016}

Our \gls{cl}-derived type-I \gls{pd} densities are in good agreement with previous results from macroscale measurements by Piva \textit{et al.}\cite{Piva2020} (Fig.\ \ref{pd_dens}), which corroborates that we are spatially resolving \textit{individual} \glspl{pd}. Furthermore, their results also associate these defects with a near-midgap energy level, and identify another energy level near the valence band edge which could match our type-II defects. Other macroscale studies have also detected two types of \glspl{pd} in InGaN/GaN \glspl{qw}.\cite{Armstrong2012, Armstrong2015} Evidence suggests midgap (type-I) defects are intrinsic nitrogen vacancy complexes,\cite{Haller2018, Han2021} while the type-II defect may be linked to impurities such as oxygen or carbon.\cite{Reshchikov2014a} All these results highlight the versatility of our \gls{cl} analysis to extract this information directly from images of \glspl{pd}, and have particular relevance for III-nitride devices based on InGaN. 

\vspace{0.5cm}

\pdfbookmark[1]{Diffusion analysis}{sec:diff}
\label{diff_sec}

\noindent\textbf{Diffusion analysis.} Carrier dynamics within \glspl{qw} is also of major relevance to optoelectronic devices. \gls{cl} has been used to gauge carrier diffusion through various methods, often requiring a specific metallic mask, reference samples, or precise knowledge of the excitation volume.\cite{Zarem1989, Jahn2006, Pauc2006, Gustafsson2010} Here we estimate diffusion lengths without having to take any new measurements, but instead from a direct analysis of the hyperspectral maps we already used to generate the images in Fig.\ \ref{pd_cl}. This allows us to investigate the impact of non-radiative \gls{pd} density on carrier diffusion, which was recently highlighted as an area where studies are lacking.\cite{Lahnemann2020}

At 10 K carriers in \mbox{InGaN/GaN} \glspl{qw} have near-zero diffusivity, and therefore negligible diffusion lengths. As the temperature is raised, the diffusion constant increases before saturating at  $\sim 200$ K.\cite{Solowan2013} The resulting increase in diffusion length reduces the spatial resolution at higher temperatures, as shown in Fig.\ \ref{pd_cl} comparing the 170 K images to those at 10 K.

\begin{figure*}[t]
	\centering
	\includegraphics[width=1\linewidth]{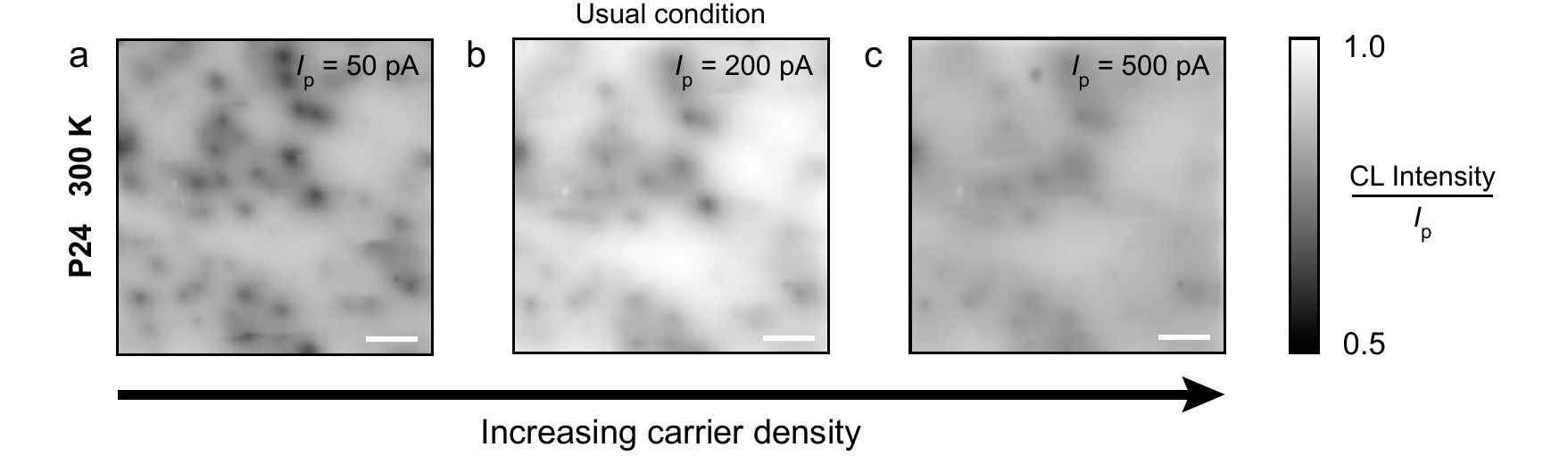}
	\caption{{\fontfamily{phv}\selectfont\footnotesize Influence of carrier density on \glspl{pd}. Large scale \glsentryshort{qw} integrated intensity images for P24 at 300 K, with the probe current ($I_{\rm p}$) varied from \textbf{a} 50 pA, to \textbf{b} 200 pA (condition used for all other results), to \textbf{c} 500 pA. \glsentryshortpl{pd} become saturated as the probe current , and hence the carrier density, is increased. All results are plotted on an intensity scale of 0.5--1, normalised relative to probe current and maximum intensity at 200 pA. Scale bars are 1 \si{\micro\metre}.}}
	\label{current}
\end{figure*}

Neglecting drift effects, carriers injected into the well diffuse from each point into 2-D Gaussian distributions, with a standard deviation, $\sigma_d$, linked to the diffusion length, $L_D$, by $L_D = \sigma_d\sqrt{2}$.\cite{Crank1975} We can therefore straightforwardly model the effect of diffusion on our \gls{cl} images by convoluting the negligibly diffused 10 K images with 2-D Gaussians. Using the convolution Gaussian's $\sigma_d$ as a fitting parameter, we can fit each 10 K image to the 170 K image of the same area, extracting $L_D$ at 170 K from the optimum fit for each sample. This method analyses average 2-D diffusion across the entire image, rather than inspecting only one local feature.

Using the \gls{cl} intensity for this procedure is not ideal, as we cannot easily account for the thermal activation and complex behaviour of non-radiative recombination at \glspl{pd}. Instead, we use the \gls{qw} peak energy images of the same areas as in Fig.\ \ref{pd_cl}, which are directly "smoothened" as the diffusion length increases. The importance of using \gls{cl} emission energy rather than intensity for diffusion analysis has been emphasised for the case of threading dislocations in GaN.\cite{Lahnemann2020} Since we are not interested in macroscopic changes in the peak energy with temperature, we subtract the mean peak energy from each image to map deviation from average peak energy, $\Delta E$. These images then reflect inhomogeneity across the \gls{qw}.

An example of this diffusion analysis being applied to P24 is shown in Figs.\ \ref{diff}a--d (see supplementary Sec.\ \ref{comp_diff_sec} for diffusion analysis of all samples). Inspecting  Fig.\ \ref{diff}c, we observe modulation in the \gls{qw} emission energy by $\sim 10$ meV with the same periodicity as the step-edges in the \gls{se} image (Fig.\ \ref{diff}a). While this modulation likely originates from the impact of step-edges on the bottom interface of the well,\cite{Tanaka1987} it is beyond the scope of this study. However, this energy modulation has a direct influence on the carrier dynamics within the \gls{qw} by presenting an energy barrier to diffusion perpendicular to the step-edges. This leads to anisotropic carrier diffusion in the \gls{qw}, which we account for by adjusting the convolution Gaussian to contain two different perpendicular standard deviations. One standard deviation was fixed parallel to the step-edge direction, resulting in a diffusion length $L_{D, \parallel}$, while the other was set perpendicular to the step-edges, giving $L_{D, \perp}$. The fitting result from convolution of the 10 K energy map is shown in Fig.\ \ref{diff}d, and compares well with the real 170 K data (Fig.\ \ref{diff}b). The computed values for the diffusion lengths confirm the anisotropic diffusion, with $L_{D, \parallel}$ being over three times larger than $L_{D, \perp}$. This step-edge induced anisotropy is likely stronger in our \glspl{qw} since the effect of interface roughness on the carrier confinement energy is magnified for thinner wells.

$L_{D, \parallel}$ and $L_{D, \perp}$ for all samples are plotted against their type-I \gls{pd} densities in the inset of Fig.\ \ref{diff}e. We can see that every sample exhibits anisotropic diffusion. Crucially, both $L_{D, \parallel}$ and $L_{D, \perp}$ are severely limited by large \gls{pd} concentrations, with $L_{D, \parallel}$ decreasing by over 80 \% from P24 to P0. In other words, despite the increased diffusion coefficient at raised temperatures, carriers in the more defective samples are unable to diffuse far since they quickly undergo non-radiative recombination at \glspl{pd}. We can link this trend to the effective lifetimes gained from \gls{trpl} (Fig.\ \ref{sample_info}d), $\tau_\textnormal{eff}$, through $L_{D} = 2 \sqrt{D \tau_\textnormal{eff}}$ where $D$ is the diffusion coefficient at 170 K. We note here that the \gls{trpl} excitation was carefully chosen to induce an initial \gls{qw} carrier density in the same order of magnitude as the carrier density in our \gls{cl} measurements ($\sim 10^{12}$ \si{\per\centi\metre\squared}, see supplementary Sec.\ \ref{carr_inj_sec}), so \gls{trpl} effective lifetimes should be comparable to carrier lifetimes in CL. The result from fitting this equation to our data is shown in Fig.\ \ref{diff}e, with extracted diffusion coefficients underscoring the highly anisotropic diffusion in our \glspl{qw}. $D_\parallel$, which represents unrestricted diffusion in the well, is in reasonable agreement with diffusion coefficients previously found for InGaN/GaN \glspl{qw}.\cite{Solowan2013} With this analysis, we have demonstrated that a straightforward and general \gls{cl} data treatment can measure diffusion lengths with only two fitting parameters and no assumed constants. Furthermore, \gls{pd} density is confirmed as a critical parameter controlling diffusion in \glspl{qw}, re-emphasising that high \gls{iqe} samples are required to study intrinsic carrier dynamics.

\vspace{0.5cm}

\pdfbookmark[1]{Influence of carrier density on point defects}{sec:carr}
\label{carr_sec}

\noindent\textbf{Influence of carrier density on point defects.}  The impact of non-radiative defects on the \gls{iqe} of \glspl{qw} is known to depend strongly on the carrier density.\cite{David2020} \gls{cl} measurements using continuous excitation produce a local steady-state carrier density, which increases with the generation rate and lifetime of the carriers. Since the generation rate is proportional to the \gls{cl} electron-beam probe current, $I_{\rm p}$,\cite{Guthrey2020} this current directly affects the carrier density in the \gls{qw}---yet $I_{\rm p}$ is not commonly varied in \gls{cl} defect studies.

In Fig.\ \ref{current}, we investigate how carrier density alters the sensitivity of \gls{cl} to \glspl{pd} by mapping the \gls{qw} intensity on the same area of P24 at three different probe currents. Due to the low density of type-I defects in P24, the images can be taken at a higher temperature and larger scale than previous results while still being able to resolve \glspl{pd}. The images are normalised relative to $I_{\rm p}$ to give an indication of how \gls{iqe} is affected. On increasing $I_{\rm p}$, the impact from \glspl{pd} is greatly reduced since defect-assisted non-radiative recombination usually dominates at lower carrier densities.\cite{David2020} As the carrier density is increased, radiative recombination and then Auger recombination progressively play larger roles. Both of these processes are unlikely to be strongly spatially correlated with defects, and hence they suppress any influence of \glspl{pd} on the \gls{cl} intensity as in Fig.\ \ref{current}c. We note that the brightness of the $I_{\rm p} = 200$ pA image (Fig.\ \ref{current}b) suggests this condition is near peak-\gls{iqe}, in agreement with the calculated steady-state \gls{cl} carrier density of $\sim10^{12}$ \si{\per\centi\metre\squared} (see supplementary Sec.\ \ref{carr_inj_sec}). 

Looking back at Figs.\ \ref{pd_cl}f--j, we can now understand why the influence of each \gls{pd} on the \gls{cl} intensity appears weaker and more "washed-out" for the samples with a low type-I \gls{pd} concentration, i.e., P16 and P24. As previously explained, the steady-state density in each sample's well scales with its carrier lifetime. Due to their longer lifetimes at high temperatures (Fig.\ \ref{sample_info}d), at 170 K the \gls{cl} carrier density in P16/P24 is higher than in the more defective samples. Consequently, this large steady-state carrier density leads to decreased sensitivity to \glspl{pd} in a similar way to what we see in Fig.\ \ref{current}. As well as directly demonstrating the dependence of \gls{pd} non-radiative recombination on the carrier density, these results emphasise that probe current, though often overlooked, must be carefully selected for optimal defect imaging in \gls{cl}.


\vspace{0.5cm}

\pdfbookmark[1]{Conclusion}{sec:conc}

In summary, combining \gls{cl} at an acceleration voltage of only 1.5 kV with specific sample design, we spatially resolved and analysed \textit{individual} non-radiative \glspl{pd} buried in \mbox{InGaN/GaN} single \glspl{qw}. Using just two hyperspectral maps of each sample at different temperatures, we extracted information on \gls{pd} types, \gls{pd} densities, and carrier diffusion lengths, all through direct observation and straightforward analysis methods. We identified two different types of \gls{pd}: near-midgap defects with a strong impact on \gls{cl} intensity and \gls{iqe} which can be incorporated into an indium-containing underlayer (type-I), and defects with a lesser impact on efficiency which are not affected by the underlayer (type-II). The density of type-I \glspl{pd} was estimated to range from $10^{14}$ \si{\per\centi\metre\cubed} up to $10^{16}$ \si{\per\centi\metre\cubed} with decreasing underlayer thickness, highlighting the capability of our methodology to image \glspl{pd} even up to high densities. With careful experimental methods, we investigated the interplay between carrier dynamics and \glspl{pd}. We measured diffusion lengths from \gls{cl} peak energy images, demonstrating their dependence on the \gls{pd} concentration, and showed that carrier density significantly affects the influence of \glspl{pd} on \gls{cl} intensity. Overall, this comprehensive study serves as a proof-of-concept for imaging and directly analysing \glspl{pd} in \glspl{qw}, and lays the foundation for advanced future studies such as applying time-resolved \gls{cl} to a single buried \gls{pd}.


\pdfbookmark[1]{Methods}{sec:meth}

\vspace{1cm}
\noindent\fontfamily{phv}\textbf{Methods}\\\normalfont\small

\noindent\textbf{Sample growth.} The samples were grown by metal-organic vapour phase epitaxy in a horizontal Aixtron 200/4 RF-S reactor on n-type \textit{c}-plane freestanding GaN substrates (dislocation density  $\sim 10^6$ \si{\per\centi\metre\squared}). After a low-temperature GaN buffer, 700 \si{\nano\metre} of GaN is deposited at 1000 \si{\celsius} using trimethylgallium with H\textsubscript{2} as the carrier gas. Then, the temperature is decreased to 770 \si{\celsius}, and the carrier gas is switched to N\textsubscript{2} for the growth of the In\textsubscript{0.17}Al\textsubscript{0.83}N \gls{sl}, using trimethylindium and trimethylaluminium. Triethylgallium is used for the GaN layer in the \gls{sl}. This lattice-matched In\textsubscript{0.17}Al\textsubscript{0.83}N/GaN \gls{sl} was used as the underlayer since it was demonstrated to produce near-ultraviolet \glspl{led} with improved \gls{iqe} compared to bulk underlayers.\cite{Haller2019a} A 10 \si{\nano\metre} GaN spacer is grown at the same temperature with triethylgallium. Up to this point, the structure is mostly grown with Si doping at around $1 \times 10^{18}$ \si{\per\centi\metre\cubed}; however, the last two periods of the \gls{sl} and the first 5 \si{\nano\metre} of the following GaN barrier are highly doped with Si ($> 1\times10^{20}$ \si{\per\centi\metre\cubed}) to screen the electric field arising from the spontaneous polarisation mismatch at InAlN/GaN interfaces (see supplementary Sec.\ \ref{tem_sec}).\cite{Haller2019a} The rest of the structure (see Fig.\ \ref{sample_info}a) is grown under similar conditions but with no intentional doping. The InGaN/GaN single \gls{qw} has a nominal indium content of 15 \%.

\vspace{0.5cm}

\noindent\textbf{Photoluminescence.} The room-temperature \gls{qw} \glsentryshort{pl} was measured for each sample using a 325 \si{\nano\metre} HeCd laser at a power density of 2.0 \si{\watt\per\centi\metre\squared}. For \gls{trpl}, we used the third harmonic of a mode-locked Ti:sapphire laser emitting 100 fs pulses at 266 \si{\nano\metre} with a repetition rate tuned to 7.6 MHz to measure extended \glsentryshort{pl} decay times. An excitation density of 5 \si{\micro\joule\per\centi\metre\squared} was applied using a 40x reflective objective. Detection was performed using a time-correlated single-photon counting system with an avalanche photo-diode. Decay curves were measured at five different energies across the \gls{qw} emission before being summed to give the final curves; this ensures any carrier transfer to different energies in the \gls{qw} has a limited impact on our measured lifetimes. Effective lifetimes were calculated by mono-exponentially fitting these decay curves at early delays accounting for the instrument response function. The decay time resolution limit of this system is around 50 \si{\pico\second}.

\vspace{0.5cm}

\noindent\textbf{Cathodoluminescence spectroscopy.} \gls{cl} imaging was performed using a \gls{cl}-dedicated scanning electron microscope system (Attolight Rosa 4634) with the acceleration voltage set at 1.5 kV. For most measurements (apart from those in Fig.\ \ref{current}) the e-beam was raster-scanned in steps of 18 \si{\nano\metre} with an integration time of 3 \si{\milli\second} and a probe current of 200 pA. A Cassegrain reflective objective was used to collect the light and direct it towards a spectrometer, with a grating of 600 lines per \si{\milli\metre} and a blaze-wavelength at around 300 nm. The dispersed light was then sent to a cooled charge-coupled device camera to record a full intensity-energy spectrum at every pixel (hyperspectral imaging). The \gls{qw} emission at every pixel was fully fitted, allowing us to generate the \gls{qw} \gls{cl} intensity and peak energy images shown in this work. The fitting function was a combination of three Gaussians to fit the main \gls{qw} peak and subsequent longitudinal-optical phonon replicas (see supplementary Sec.\ \ref{hyp_fit_sec}). Temperature control was achieved using an open-flow helium cryostat. \gls{se} images taken at different temperatures on the same sample area were precisely aligned by finding the relative pixel shifts which maximised the cross-correlation value---applying these same shifts to the simultaneously-acquired hyperspectral maps then spatially aligned the \gls{cl} data. All hyperspectral analysis was carried out using the \textit{hyperspy} package in Python.\cite{Pena2020}

\vspace{0.5cm}

\noindent\textbf{Data availability.} The datasets generated during and/or analysed during the current study are available from the corresponding author on reasonable request.

\vspace{0.5cm}
\noindent\fontfamily{phv}\textbf{Acknowledgements}\\\normalfont
This work was supported by the Swiss National Science Foundation through Grant No.\ 200021E 175652. The CIME at EPFL is acknowledged for access to its facilities, and Dr.\ B.\ Bártová at CIME is thanked for the focused ion beam preparation. We thank N.\ Tappy (EPFL), Dr.\ C.\ Haller (EPFL), and Prof.\ C.\ Hébert (EPFL), for useful discussion. 

\vspace{0.5cm}

\noindent\fontfamily{phv}\textbf{Author contributions}\\\normalfont
J.-F.C. grew the samples. T.F.K.W. performed the room-temperature \gls{pl} measurements, while T.F.K.W. and W.L. carried out the \gls{cl} measurements. V.O. performed all \gls{trpl} measurements with R.A.T. D.T.L.A. carried out the scanning transmission electron microscopy characterisation. T.F.K.W. was responsible for all analysis carried out on the raw data from the above measurements. T.F.K.W., R.B., and N.G. wrote the paper. N.G. conceived and directed the project. All authors discussed the results and commented on the manuscript.

\vspace{0.5cm}
\noindent\fontfamily{phv}\textbf{Additional information}\\\normalfont
\textbf{Supplementary Information} accompanies this paper. \\
\textbf{Competing interests:} The authors declare no competing financial interests.

\onecolumngrid
\clearpage
\large

\pdfbookmark[0]{Supplementary information}{sec:suppl}

\begin{center}
	\textbf{\Large SUPPLEMENTARY INFORMATION}
\end{center}

\setcounter{equation}{0}
\setcounter{figure}{0}
\setcounter{table}{0}
\makeatletter
\renewcommand{\theequation}{S\arabic{equation}}
\renewcommand{\thefigure}{S\arabic{figure}}
\renewcommand{\thetable}{S\arabic{table}}
\newcounter{SIfig}
\renewcommand{\theSIfig}{S\arabic{SIfig}}

\section{CATHODOLUMINESCENCE HYPERSPECTRAL FITTING}
\label{hyp_fit_sec}

\begin{figure}[h]
	\centering
	\includegraphics[width=1\columnwidth]{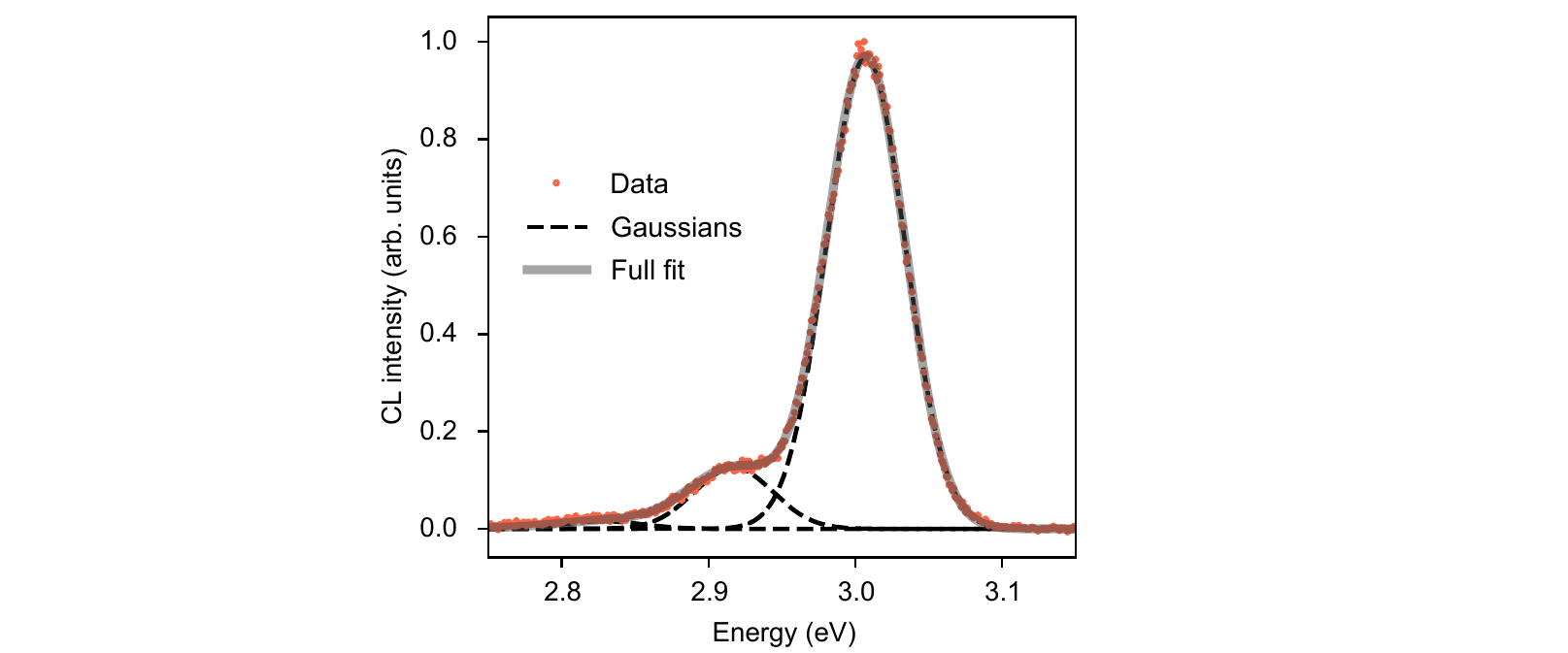}
	\caption{ \normalsize Example of a measured \glsentryshort{cl} spectrum acquired at one pixel in a hyperspectral map of sample P16 at 10 K, along with its respective best fit from the triple-Gaussian model.}
	\refstepcounter{SIfig}\label{fitting}
\end{figure}

As mentioned in the main text, \gls{qw} \gls{cl} integrated intensity and peak energy maps were obtained from hyperspectral maps by fitting the \gls{cl} spectrum at each pixel. All fitting was done using the \textit{hyperspy} package for Python.\cite{Pena2020} An example \gls{cl} spectrum taken from a hyperspectral map of P16 is shown in Fig.\ \ref{fitting}. The emission shape is very characteristic for InGaN/GaN \glspl{qw} with significant inhomogeneous broadening and clear \gls{lo} phonon replicas at lower energies. Due to the large inhomogeneous broadening, the spectra are well fitted by using three Gaussians: one for the main peak, and two for the subsequent \gls{lo} phonon replicas. To minimise free parameters, we fixed the peak energy separation of these three Gaussians at 92 meV to agree with the \gls{lo} phonon energy in wurtzite GaN.\cite{Harima2002} The peak intensities of the phonon replicas were fixed to that of the main peak using the Huang-Rhys factor as a fitting parameter.\cite{Zhang2001} Finally, the \glspl{fwhm} of the phonon replicas were fixed to be equal---this left the model with only five free fitting parameters. The best fit using this model is shown in Fig.\ \ref{fitting}, with each individual Gaussian component also plotted.

With accurate heteroscedastic noise estimation, we obtained reduced chi-squared ($\chi^2_R$) values for every spectrum fitted. All hyperspectral maps fitted in this study had an average $\chi^2_R$ in the range 0.9--1.6, confirming the suitability of our model. 

\section{CARRIER INJECTION IN CATHODOLUMINESCENCE}
\label{carr_inj_sec}

\begin{figure}[h]
	\centering
	\includegraphics[width=1\columnwidth]{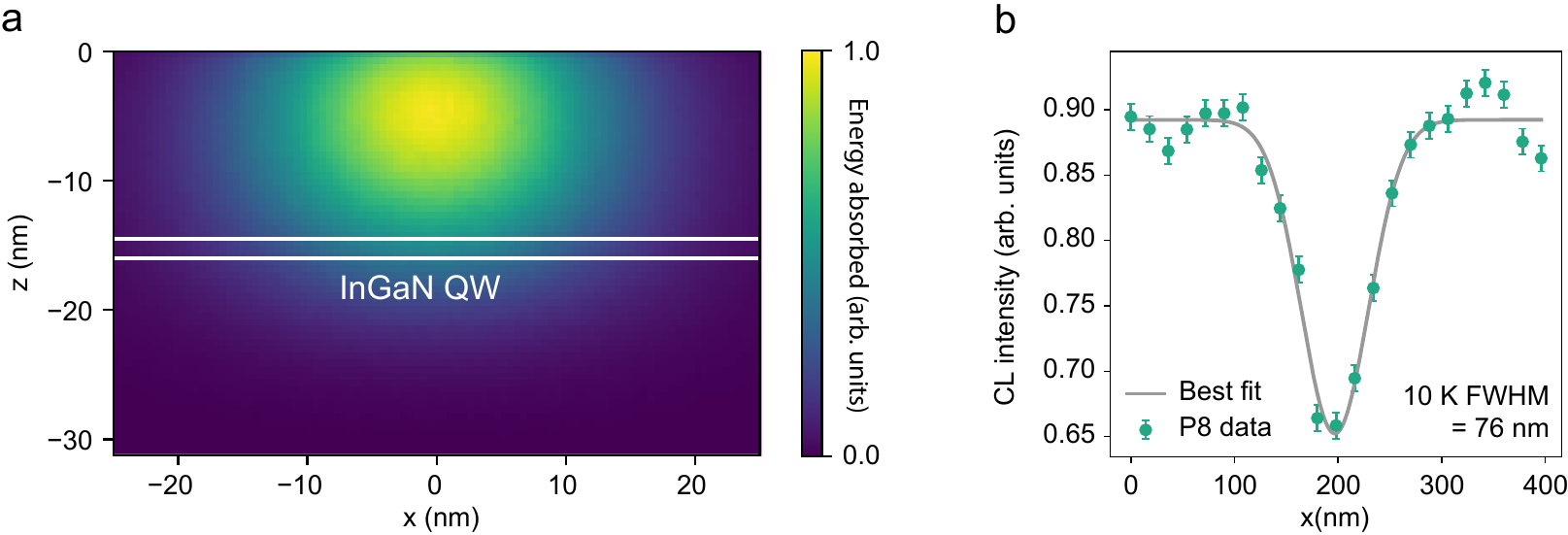}
	\caption{ \normalsize \textbf{a} Interaction volume at 1.5 kV calculated by Monte-Carlo simulation (using \textit{Casino})\protect\cite{Drouin2007}, with the position of the InGaN/GaN \glsentryshort{qw} highlighted. Energy absorbed is in units of energy per unit volume, normalised to the maximum value. \textbf{b} Experimental \glsentryshort{cl} \glsentryshort{qw} intensity profile measured across a single \glsentryshort{pd} at 10 K in P8. The profile is fitted with a Gaussian, resulting in a \glsentryshort{fwhm} of 76 nm.}
	\refstepcounter{SIfig}\label{injection}
\end{figure}

Fig.\ \ref{injection}a shows the interaction volume of the electron beam at an acceleration voltage of 1.5 kV, gained from a Monte-Carlo simulation. For this simulation, we estimated the electron probe diameter at the sample surface by inspecting the \gls{se} images in Fig.\ \ref{pd_cl} of the main text. Specifically, we set the \gls{fwhm} of the probe to the average \gls{fwhm} of surface step-edges ($\sim25$ nm), since these step-edges should be near-atomically sharp. Although this is a conservative estimate, the probe diameter is expected to be larger at low acceleration voltages due to enhanced chromatic aberration, particularly in a system lacking beam deceleration such as ours. The potential loss in \gls{cl} signal resolution due to the increased probe diameter compared to higher acceleration voltages is negligible given the reduction in interaction volume this low acceleration voltage provides. 

From Fig.\ \ref{injection}a, we can see that the vast majority of the excitation energy (95 \%) is absorbed in the top 20 nm of the sample, hence the InGaN/GaN single \gls{qw} is placed only 15 nm below the surface. This design ensures all generated carriers are able to reach the \gls{qw} in the entire temperature range we explore, since the carrier diffusion length in GaN is always above 15 nm from 10 K up to 300 K.\cite{Kaganer2019, Brandt2020, Lahnemann2020} In addition, using thin top layers limits any lateral carrier spreading that can occur before the carriers relax to the \gls{qw}. The radial \gls{fwhm} of the interaction volume in Fig.\ \ref{injection}a is around 25 nm, which indicates the maximum achievable \gls{cl} resolution if there were no lateral carrier diffusion. Thus we could expect this to match the \gls{cl} resolution of our 10 K images, since the diffusion length should be near-zero in the \gls{qw} as emphasised in the main text.

To compare the results of the Monte-Carlo simulation to our experiments, we inspect the 10 K \gls{qw} \gls{cl} intensity across one \gls{pd} as shown in Fig.\ \ref{injection}b (taken from Fig.\ \ref{pd_cl}m in the main text). Since the \gls{pd} itself is localised on one point in the \gls{qw}, the \gls{fwhm} of this profile indicates the real resolution in the \gls{qw} at 10 K. The experimentally measured resolution, at 76 nm, is larger than the 25 nm that was expected from Fig. \ref{injection}a. This discrepancy is confirmed by repeating the analysis for different \glspl{pd}, with an average 10 K intensity profile \gls{fwhm} of $80 \pm 10$ nm. 

The mismatch is explained by two factors. Firstly, a recent study by Jahn \textit{et al.}\ has evidenced an increase in \gls{cl} interaction volumes compared to Monte-Carlo simulations,\cite{Jahn2020} since \gls{cl} generates hot carriers which can only radiatively recombine once they have lost excess energy through phonon emission. They well-described the interaction volume broadening in GaN by convoluting the Monte-Carlo radial profile by a Gaussian with \gls{fwhm} of about 50 nm. Secondly, as previously mentioned, there will be some lateral carrier diffusion in the top barrier of the sample before carriers fully relax to the \gls{qw}. With random diffusion, we expect this effect to broaden the final 2-D carrier distribution in the \gls{qw} on the order of the thickness of the barrier, i.e., 15 nm. Combining both of these factors with our original 25 nm Monte-Carlo \gls{fwhm}, we arrive at a more realistic resolution estimate of $\sim60$ nm, much closer to our measured value of $80 \pm 10$ nm. The remaining difference of 10--30 nm could then be due to the limited diffusion possible within the \gls{qw} at 10 K, with a diffusion coefficient of far less than 0.01 \si{\centi\metre\squared\per\second}.\cite{Solowan2013} Such low levels of diffusion at 10 K would only slightly impact our calculated \gls{qw} diffusion lengths at 170 K, since these are mainly on the order of 100s of nanometres.

The $80 \pm 10$ nm \gls{fwhm} of the \gls{cl} intensity profile around single \glspl{pd} should also be close to the \gls{fwhm} of the carrier density distribution at 10 K in the \gls{qw}. Meanwhile we can estimate the total generation rate, $G_{\textnormal{tot}}$ (\si{\per\second}), of carriers in \gls{cl} through the well known equation:\cite{Guthrey2020}

\begin{equation}
G_{\textnormal{tot}} = \frac{I_{\rm p}}{q} . \frac{E_{\textnormal{dep}}}{3 E_{\rm g}} \hspace{3mm},
\end{equation}

\noindent where $I_{\rm p}$ is the electron beam probe current, $q$ is the charge of an electron, $E_{\rm g}$ is the bandgap of the sample, and $E_{\textnormal{dep}}$ is the average energy deposited per electron in the sample, which is equivalent to the beam energy minus the energy lost through backscattered electrons (calculated from Monte-Carlo simulations).\cite{Drouin2007} Using our estimated steady-state carrier distribution \gls{fwhm}, $\Gamma$, and assuming this distribution is Gaussian-like, we can then write the peak steady-state carrier density in the \gls{qw},  $n_{\textnormal{QW}}$, as:

\begin{equation}
n_{\textnormal{QW}} =  G_{\textnormal{tot}} \tau. \frac{4 \ln(2)}{\pi \Gamma^2} = \frac{I_{\rm p}}{q} . \frac{E_{\textnormal{dep}}}{3 E_{\rm g}}  \tau. \frac{4 \ln(2)}{\pi \Gamma^2} \hspace{3mm},
\label{nss}
\end{equation}

\noindent if we assume all generated carriers relaxed to the \gls{qw}, with $\tau$ being the carrier lifetime. Into eq.\ \ref{nss} we input $I_{\rm p} = 200$ pA, $E_{\rm g} = E_{g, \textnormal{GaN}} = 3.50$ eV, $\Gamma = 80 \pm 10$ nm, and the calculated $E_{\textnormal{dep}} = 1.12$ keV. For $\tau$, we use the effective lifetime gained from \gls{trpl} results at 10 K ( $\sim 1.5$ ns, see \hyperref[trpl_sec]{next section}). The resulting $n_{\textnormal{QW}}$ is found to be $2.8 \pm 0.5 \times 10^{12}$ \si{\per\centi\metre\squared}. This calculated density is important when considering the comparison of \gls{cl} diffusion lengths and \gls{trpl} effective lifetimes in Fig.\ \ref{diff} of the main text, as described in the \hyperref[trpl_sec]{next section}. We note that peak \gls{iqe} for high-quality \mbox{InGaN/GaN} \glspl{qw} commonly occurs at about this carrier density,\cite{David2020} which is in good agreement with what we observe in Fig.\ \ref{current} of the main text, in which the highest \gls{cl} intensity-probe current ratio is achieved at $I_{\rm p} = 200$ pA. Conversely, at $I_{\rm p} = 50$ pA we expect a carrier density in the range $10^{11}$ \si{\per\centi\metre\squared} while for $I_{\rm p} = 500$ pA it should be closer to $10^{13}$ \si{\per\centi\metre\squared}---these densities are further into the defect dominated and Auger recombination regimes, respectively.\cite{David2020}

\section{TIME-RESOLVED PHOTOLUMINESCENCE}
\label{trpl_sec}

\begin{figure}[h]
	\centering
	\includegraphics[width=1\columnwidth]{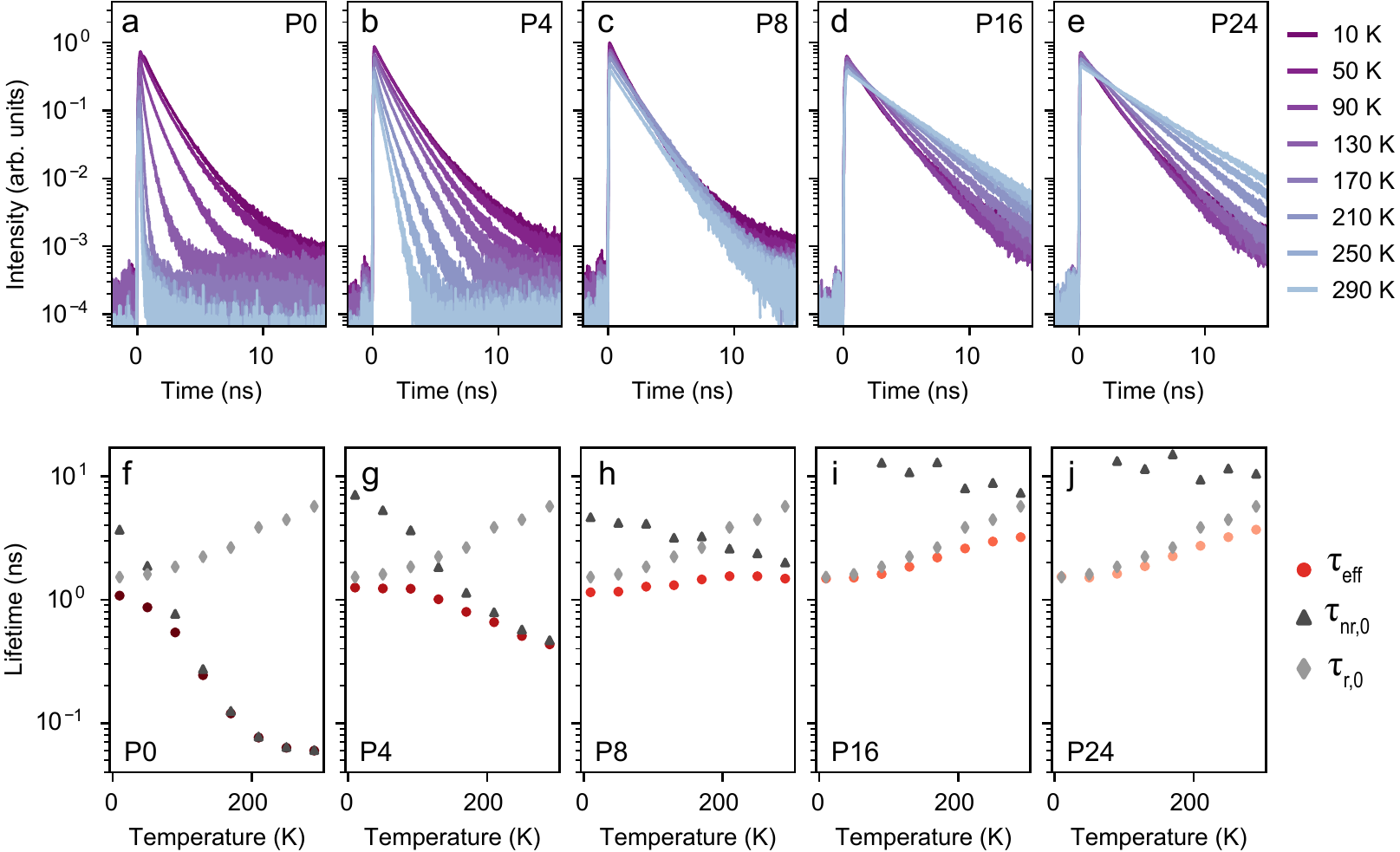}
	\caption{ \normalsize \textbf{a--e} Temperature-dependent \glsentryshort{trpl} decay curves of P0, P4, P8, P16, and P24, respectively. $\tau_{\textnormal{eff}}$ are extracted from mono-exponential fitting of \glsentryshort{trpl} intensities at early delays, displayed in \textbf{f--j} for all samples along with calculated $\tau_{\rm r, 0}$ and $\tau_{\rm nr, 0}$ values.}
	\refstepcounter{SIfig}\label{trpl_fig}
\end{figure}

In Fig.\ \ref{sample_info} of the main text, data extracted from \gls{qw} \gls{trpl} decay curves are presented: here we show the original curves in Figs.\ \ref{trpl_fig}a--e. The evolution of PL intensity, $I$, with time, $t$, in these decay curves is described by the function:

\begin{equation}
I = A \frac{n_0}{\tau_{\rm r}} \exp\bigg(\frac{-t}{\tau}\bigg) \hspace{3mm},
\label{overall_trpl}
\end{equation}

\noindent where $n_0$ is the initial carrier density in the well, $\tau_{\rm r}$ is the carrier radiative lifetime, $\tau$ is the total carrier lifetime, and $A$ is a constant related to the collection efficiency, area of sample excited, and integration time. $\tau$ is related to $\tau_{\rm r}$ and the non-radiative lifetime, $\tau_{\rm nr}$, through:

\begin{equation}
\frac{1}{\tau} = \frac{1}{\tau_{\rm r}} + \frac{1}{\tau_{\rm nr}} \hspace{3mm}.
\label{tot_lifetime}
\end{equation}

Unfortunately, $\tau_{\rm r}$ and $\tau_{\rm nr}$ often depend on carrier density in a non-straightforward manner.\cite{Liu2016, David2020} Since the carrier density in \gls{trpl} decreases as time progresses, this dependence leads to deviation from mono-exponential decay as we see in Figs.\ \ref{trpl_fig}a--e. To simplify our analysis, we obtained the effective lifetime, $\tau_{\textnormal{eff}}$, for each curve by fitting intensity at early delay times with a mono-exponential decay convoluted with the instrument response function. $\tau_{\textnormal{eff}}$ is then equivalent to the true carrier lifetime, $\tau$, only when the carrier density is around $n_0$. With an excitation density of 5 \si{\micro\joule\per\centi\metre\squared} at a laser wavelength of 266 nm, and making the simplifying assumption that nearly all carriers relax to the \gls{qw}, $n_0$ is on the order of $10^{12}$ \si{\per\centi\metre\squared}. Crucially, this is the same order of magnitude as the steady-state carrier densities in our \gls{cl} measurements (see \hyperref[carr_inj_sec]{previous section}), allowing us to use our calculated $\tau_{\textnormal{eff}}$ as a reasonable approximation for the carrier lifetime in \gls{cl}, as we did in the \hyperref[diff_sec]{diffusion analysis} of the main text and in the \hyperref[carr_inj_sec]{previous section} of this supplementary. As mentioned previously, this carrier density should also be near peak-\gls{iqe} conditions for high-quality \mbox{InGaN/GaN} \glspl{qw}.\cite{Liu2016, David2020}

To calculate the macroscopic \glspl{iqe} as presented in Fig.\ \ref{sample_info}e of the main text, we used a method first applied by Langer \textit{et al}.\cite{Langer2013} Eq.\ \ref{overall_trpl} indicates that using the initial intensity of a decay curve, $I_0$, we can calculate the initial radiative lifetime, $\tau_{\rm r, 0}$, at any temperature, $T$, through:

\begin{equation}
\tau_{\rm r, 0}(T) = \frac{A n_0}{I_0(T)} \hspace{3mm}.
\end{equation}

\noindent Since excitation and detection conditions aren't changed, $A$ and $n_0$ will be near-constant across all temperatures as long as the intensity rise time is much less than $\tau_{\textnormal{eff}}$. Hence, if we have $\tau_{\rm r, 0}$ at one temperature, we can compute $A n_0$, which then lets us calculate $\tau_{\rm r, 0}$ for all temperatures. Having already found $\tau_{\textnormal{eff}}$, we can then determine $\tau_{\rm nr, 0}$ through eq.\ \ref{tot_lifetime}. Finally, the \gls{iqe} is given by:

\begin{equation}
\textnormal{IQE}(T) =\frac{\tau_{\textnormal{eff}}(T)}{\tau_{\rm r, 0}(T)} \hspace{3mm}.
\label{iqe_eq}
\end{equation}

The next step is to estimate $\tau_{\rm r, 0}$ at one temperature. To this end, we look more closely at the temperature-dependence of P24 $\tau_{\textnormal{eff}}$ (Fig.\ \ref{trpl_fig}j). The lifetime of P24 increases linearly above $\sim$ 90 K from 1.6 \si{\nano\second} up to 3.7 \si{\nano\second} at room temperature. Such behaviour is expected for $\tau_{\rm r}$ in any direct bandgap semiconductor \gls{qw} for both free electron/hole pairs and excitons.\cite{Matsusue1987, Andreani1991} Meanwhile below 90 K, the lifetime is temperature-independent, indicating that the majority of carriers are localised in this temperature range due to alloy disorder in the \mbox{InGaN/GaN} \gls{qw}.\cite{Feldmann1987, Hangleiter2017} The consistency of P24's $\tau_{\textnormal{eff}}$ behaviour with that expected for $\tau_{\rm r}$ indicates this sample is dominated by radiative recombination across the full temperature range we explore. Given that defect-assisted non-radiative recombination is a thermally-activated process,\cite{Henry1977} we can safely assume that $\tau_{\textnormal{eff}}$ for P24 is very close to $\tau_{\rm r, 0}$ at 10 K ($\sim 1.5$ ns). We use this value as $\tau_{\rm r, 0}(T = 10 \textnormal{ K})$ for all samples, since they are expected to have the same radiative lifetime behaviour due to identically grown \gls{qw} structures.

With this key, we can now calculate $\tau_{\rm r, 0}(T)$ for all samples using the process described above. The main source of random error in this method derives from small changes in optical alignment altering the value of $A n_0$. We minimise the error by averaging $\tau_{\rm r, 0}(T)$ across the samples, which is reasonable since they should all have similar radiative lifetimes. P0 is excluded from this average since at high temperatures its $\tau_{\textnormal{eff}}$ becomes very short (Fig.\ \ref{trpl_fig}f). This leads to highly inaccurate $\tau_{\rm r, 0}$ estimation since (i) $\tau_{\textnormal{eff}}$ is on the order of the rise time, significantly decreasing $n_0$ from its low-temperature value, and (ii) $\tau_{\textnormal{eff}}$ reaches the resolution limit of the detection system, and is therefore an overestimation of the true lifetime. 

All measured and calculated lifetimes are shown in Figs.\ \ref{trpl_fig}f--j. $\tau_{\rm nr, 0}$ is shorter at high temperatures and for thinner underlayer samples as expected. With these data, we calculated \gls{iqe}(T) with eq.\ \ref{iqe_eq} as displayed in Fig.\ \ref{sample_info}e of the main text.

\section{ATOMIC FORCE MICROSCOPY}
\label{afm_sec}

\begin{figure}[h]
	\centering
	\includegraphics[width=1\columnwidth]{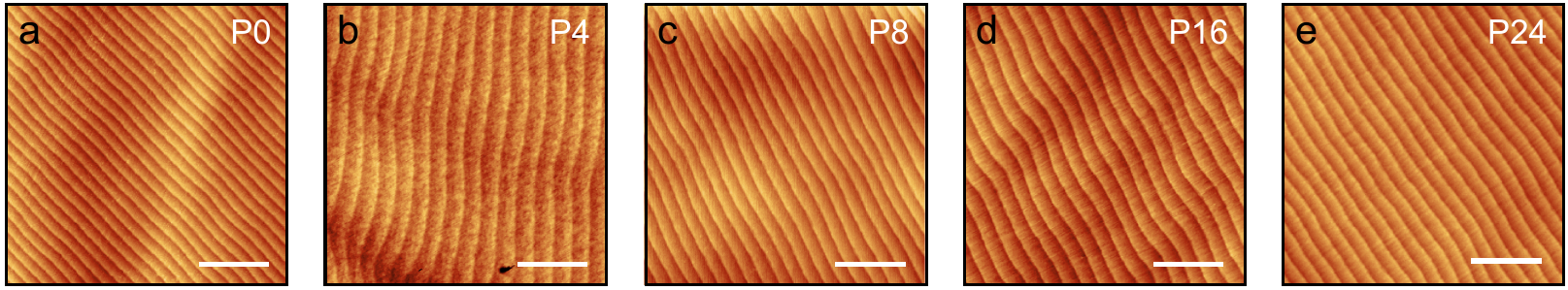}
	\caption{ \normalsize \textbf{a--e} \glsentryshort{afm} images of the surface for P0, P4, P8, P16, and P24, respectively (scale bars 500 nm, $\Delta z = 2$ nm). All samples present an evenly spaced step-flow structure.}
	\refstepcounter{SIfig}\label{afm_fig}
\end{figure}

Fig.\ \ref{afm_fig} presents complementary \gls{afm} images for all samples in this study. The observed morphology matches that of the \gls{se} images in Figs.\ \ref{pd_cl}a--e of the main text, with step-edges evenly spaced by about 80--100 nm. The step-edge height is $\sim 0.52$ nm, equivalent to two molecular monolayers of GaN.\cite{Vurgaftman2003} This places the substrate misorientation in the range 0.3$^{\circ}$--0.4$^{\circ}$, close to the nominal misorientation of 0.2$^{\circ}$. \gls{afm} underscores the low density of threading dislocations in these samples, with no surface V-pits visible at this scale of $2\times2$ \si{\micro\metre\squared}.

\section{POINT DEFECT COUNTING DETAILS}
\label{pd_sec}

\begin{figure}[b!]
	\centering
	\includegraphics[width=1\columnwidth]{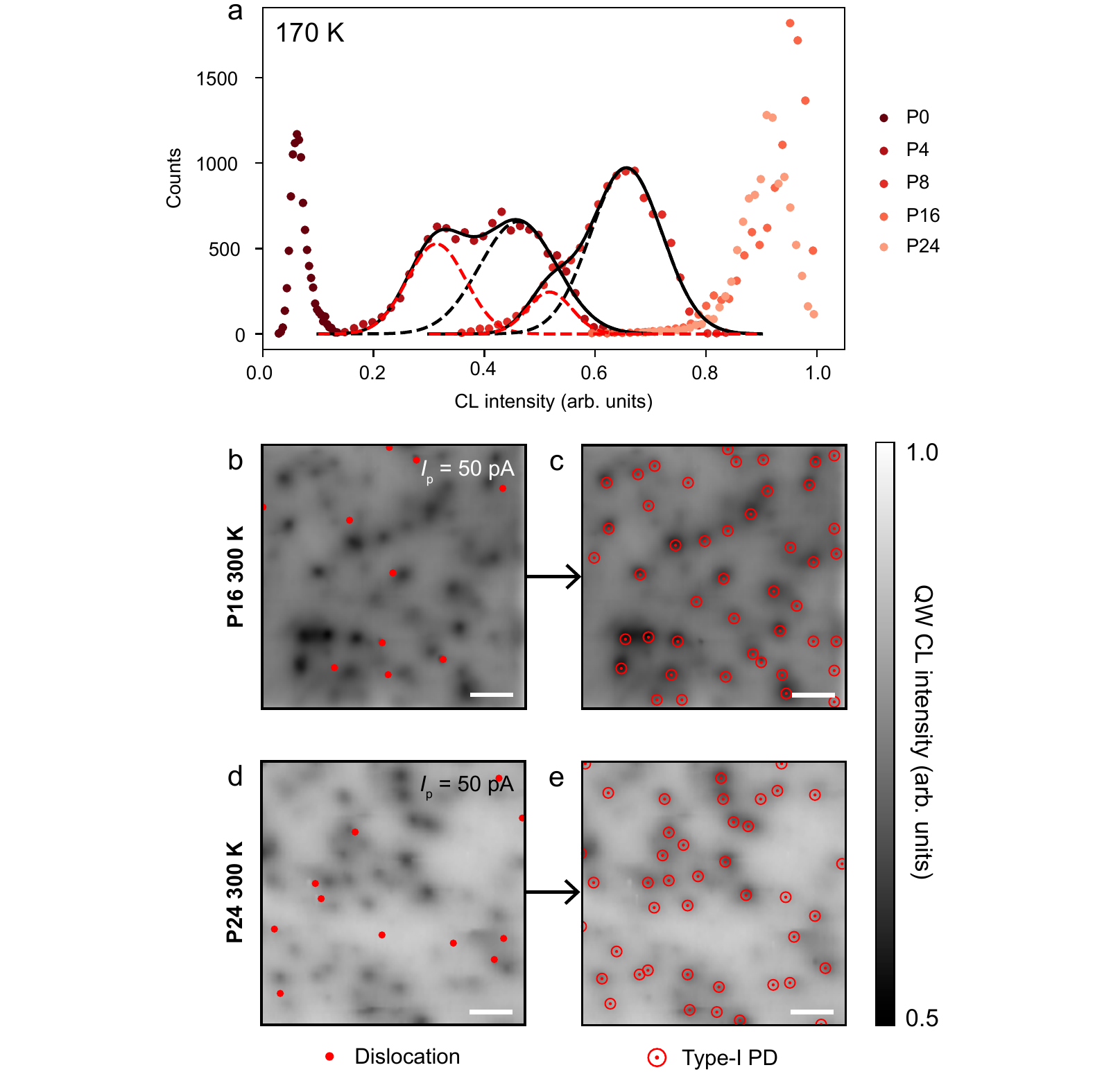}
	\caption{ \normalsize \textbf{a} Histograms of the 170 K \glsentryshort{qw} \glsentryshort{cl} intensity images shown in Figs.\ \ref{pd_cl}f--j of the main text. The results for P4 and P8 have been fitted with two Gaussian distributions to separate the contribution from type-I and type-II defects---the lower intensity Gaussian (red dashed line) peak position is taken as $I_\textnormal{lim}$ for type-I \glsentryshort{pd} counting. Large-scale integrated \gls{cl} intensity images acquired at 300 K with $I_{\rm p}=50$ pA are required to count type-I \glsentryshortpl{pd} for \textbf{b} P16 and \textbf{d} P24. The positions of threading dislocations are highlighted by red dots to show the lack of correlation with low \glsentryshort{cl} intensity areas. These images are repeated with detected type-I \glsentryshortpl{pd} indicated by red targets in \textbf{c} and \textbf{e}. (All scale bars are 1 \si{\micro\metre}).}
	\refstepcounter{SIfig}\label{pd_method}
\end{figure}

The \gls{pd} identification and counting procedure used in the main text involves three parameters: the \gls{fwhm} of the Gaussian filter ($\Gamma_{\rm G}$) and the detection threshold ($I_\textnormal{det}$) involved in the \gls{log} method applied at 10 K, along with the intensity limit ($I_{\textnormal{lim}}$) defined for the 170 \si{\kelvin} data. $\Gamma_{\rm G}$ is linked to the size of the dark regions that will be detected by the \gls{log} method. $I_\textnormal{det}$ was always fixed at 0.025 to avoid detecting random noise in the \gls{cl} images as defects.

To define a reasonable $I_\textnormal{lim}$ for each sample, we calculated the 170 \si{\kelvin} \gls{cl} intensity histograms from the $2 \times 2$ \si{\micro\metre\squared} images in Figs.\ \ref{pd_cl}f--j of the main text. P4 and P8 show distinct double-humped distributions (Fig.\ \ref{pd_method}a), as is expected from the influence of two different types of \gls{pd}. Fitting both distributions with two Gaussians then separates the contribution from type-I and type-II defects. The peaks of the lower intensity Gaussians are taken to define $I_\textnormal{lim}$ for each sample, in accordance with the greater impact of type-I defects on 170 \si{\kelvin} \gls{cl} intensity.

The other samples do not exhibit double-peaked distributions since they are dominated by only one type of defect. P0 exhibits a distribution skewed to low intensities, due to the high density of type-I \glspl{pd} overwhelming any impact from type-II \glspl{pd}. In this case, we make the simplifying assumption that all 10 K intensity fluctuations detected by \gls{log} correspond to type-I defects for P0. The results in Fig.\ \ref{pd_dens} of the main text for the P0/P4/P8 type-I and type-II \gls{pd} densities were obtained by applying our method to the $2 \times 2$ \si{\micro\meter\squared} images in Fig.\ \ref{pd_cl} of the main text, with the errors estimated by varying $\Gamma_{\rm G}$ in the range $70\leq\Gamma_{\rm G}\leq90$ nm---this range matches the \gls{cl} resolution at 10 K ($80 \pm 10$ nm, see Sec.\ \ref{carr_inj_sec}). 

On the other hand, the P16/P24 170 K images have distributions skewed towards high intensity, since (i) there are nearly no type-I \glspl{pd} at this $2 \times 2$ \si{\micro\metre\squared} scale and (ii) the higher carrier density in these samples minimises the impact of \glspl{pd} (see Fig.\ \ref{current} in the main text). To counteract both of these effects, we analysed larger $7 \times 7$ \si{\micro\metre\squared} \gls{cl} intensity images of P16/P24 at a low probe current, $I_{\rm p}$, of 50 pA (Figs.\ \ref{pd_method}b--e). We further enhanced the impact of type-I \glspl{pd} by heating the samples to room temperature. At this temperature type-II \glspl{pd} will have no resolvable effect on the intensity, so dark areas will correspond to type-I defects and we do not need to define $I_\textnormal{lim}$. This option was not available for the other samples since the longer diffusion length at high temperature combined with the high density of type-I \glspl{pd} would have made individual defects unresolvable.

At this large scale, there are more threading dislocations in the analysed area which we must not falsely identify as \glspl{pd}, hence we highlight dislocation positions in Figs.\ \ref{pd_method}b \& d (as linked to V-pits in the corresponding \gls{se} images). There is no clear spatial correlation between the dislocation positions and any dark spots in the \gls{cl} intensity. This emphasises the dominant role played by \glspl{pd} in these \glspl{qw} compared to dislocations, and allows us to safely apply \gls{log} detection to these images to calculate type-I \gls{pd} density without counting any dislocations (Figs.\ \ref{pd_method}c \& e). For error estimation, $\Gamma_{\rm G}$ was varied in the range $250\leq\Gamma_{\rm G}\leq350$ nm accounting for the much greater diffusion length at higher temperatures (c.f., P24 has diffusion length $254 \pm 10$ nm at 170 K, see Fig.\ \ref{diff} in the main text). Taking the difference between this type-I density and the density of 10 \si{\kelvin} dark areas in Figs.\ \ref{pd_cl}n \& o of the main text then allowed us to calculate the P16/P24 type-II \gls{pd} densities. 

The final number of type-I and type-II \glspl{pd} counted in this work, and the area they were counted over, is listed for each sample in Table \ref{pd_count}. We note that the low uncertainty for the P8 type-I density is a consequence of the low number of defects counted for this sample, and is very likely an underestimation.

\begin{table}[h!]
	\caption{ \normalsize \label{pd_count}Number of type-I and type-II \glspl{pd} counted for each sample in this work, along with the area these point defects were counted within; these values were used to calculate the \gls{pd} densities for each sample. The uncertainties arise from varying the $\Gamma_{\rm G}$ value used in the \gls{log} dark region detection method.}
	\begin{ruledtabular}
		\begin{tabular}{lccr}
			Sample & Type-I \gls{pd} counted & Type-II \gls{pd} counted & Area analysed (\si{\micro\metre\squared})\\
			P0 & $151 \pm 41$ & 0 (No $I_{\rm lim}$) & 4 \\
			P4 & $39 \pm 5$ & $71 \pm 21$ & 4\\
			P8 & $12 \pm 1$ & $70 \pm 16$ & 4\\
			P16 & $68 \pm 10$ & $54 \pm 2$ & 49 (type-I); 4 (type-II)\\
			P24 & $50 \pm 6$ & $58 \pm 4$ & 49 (type-I); 4 (type-II)\\
		\end{tabular}
	\end{ruledtabular}
\end{table}

\section{TRANSMISSION ELECTRON MICROSCOPY AND TRANSITION ENERGY SIMULATIONS}
\label{tem_sec}

To confirm the exact structure of our three-\gls{ml} \glspl{qw}, samples P0 and P24 were characterised by \gls{stem} \gls{eds}. The cross-section samples for \gls{stem} analysis were prepared by focused Ga ion beam lift-out using a Zeiss NVision 40. Before milling, the top surface was protected by depositing an amorphous carbon layer using an electron beam and then the ion beam. The samples were sectioned within a few degrees of the $\langle 1 1 \bar{2} 0\rangle$ zone axis. Primary milling was made using a 30 kV ion beam, with final cleaning made using a 5 kV ion beam. The atomic resolution \gls{stem} imaging and \gls{eds} spectroscopy were done using a double-aberration corrected FEI Titan Themis 60-300, using a convergence semi-angle of 20 \si{\milli\radian} and a high tension of 80 kV in order to prevent electron-beam damage by the knock-on mechanism.\cite{Smeeton2003, Baloch2013} A Fischione photomultiplier tube detector was used for taking the high-angle annular dark-field images, using an inner collection semi-angle of ~50 \si{\milli\radian}. The resolution of the images is $\sim$1.25--1.5 \si{\angstrom}. \Gls{eds} hyperspectral data were acquired using FEI Super-X ChemiSTEM detectors with FEI/Thermo Scientific Velox software; this software was also used to record the \gls{stem} images and for making the \gls{eds} data analysis. In order to gauge the thickness of the \gls{stem} samples, and as a check on \gls{eds} results, complementary \gls{stem} electron energy-loss spectra were acquired using a Gatan GIF Quantum ERS spectrometer: the measured thicknesses were 70 \si{\nano\metre} and 40 \si{\nano\metre} for P0 and P24, respectively.

\begin{figure}[h]
	\centering
	\includegraphics[width=1\columnwidth]{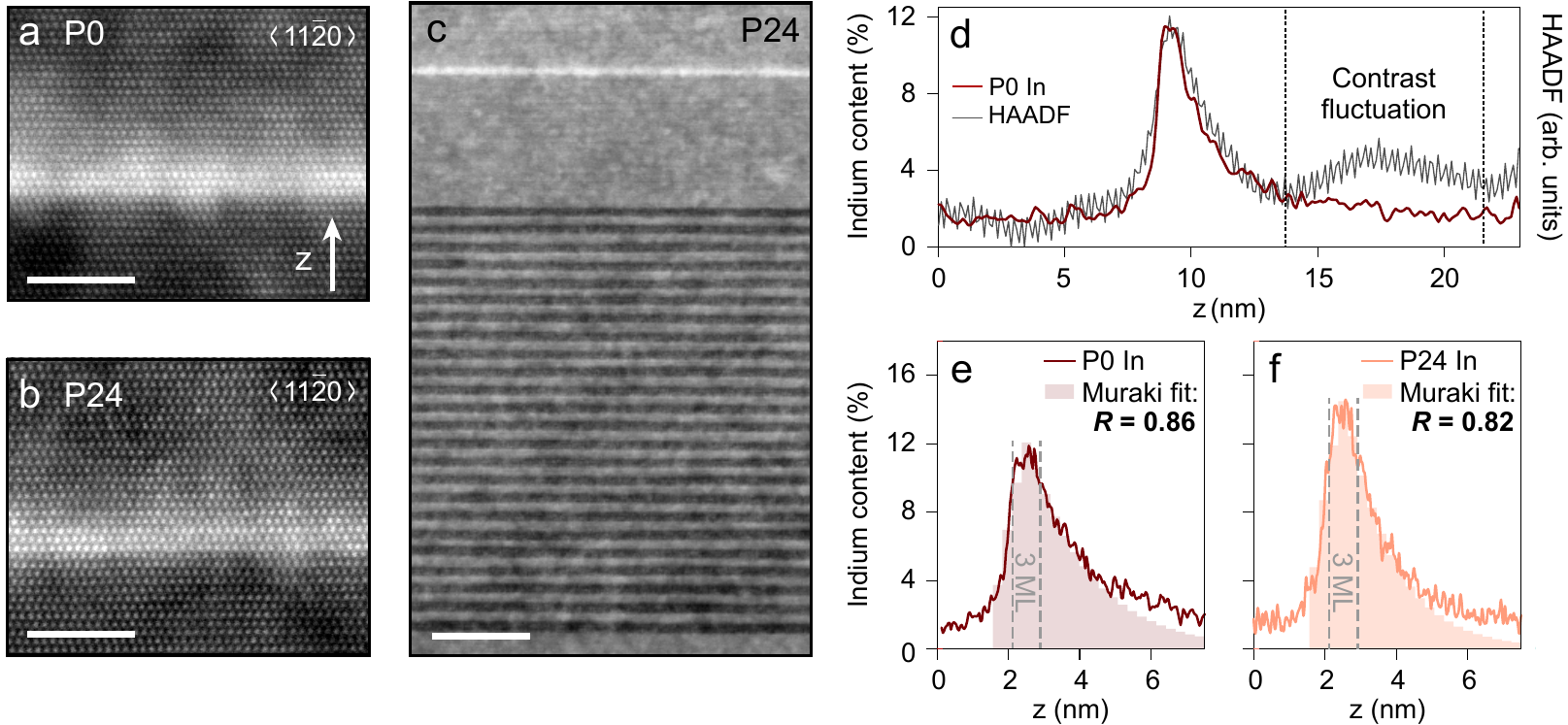}
	\caption{ \normalsize \glsentryshort{stem}-\glsentryshort{haadf} images of the \gls{qw} in \textbf{a} P0 and \textbf{b} P24, both taken down the $\langle 11\bar20\rangle$ axis (scale bars 2 nm). \textbf{c} \glsentryshort{stem}-\glsentryshort{haadf} image of the P24 sample structure down the $\langle 11\bar20\rangle$ axis, showing the InAlN/GaN superlattice and the InGaN/GaN single \glsentryshort{qw} (scale bar 20 nm). \textbf{d} \glsentryshort{haadf} contrast profile across the QW of P0 compared with the indium content profile extracted from the same region using \glsentryshort{eds}. Dashed lines mark the boundary of an artefactual contrast fluctuation unrelated to indium. Indium profiles across the \gls{qw} ($z$ axis) for \textbf{e} P0 and \textbf{f} P24 from \gls{eds}. The Muraki fit is shown as a shaded area.}
	\refstepcounter{SIfig}\label{tem_fig}
\end{figure}

Fig.\ \ref{tem_fig} presents the results, starting with atomic-resolution \gls{haadf} images of the P0 and P24 \glspl{qw} [Figs.\ \ref{tem_fig}a \& b]. The \glspl{qw} are close to the nominal thickness of three \glspl{ml}, but their interfaces are not completely sharp. However, this observation may be affected by contrast fluctuations which are unrelated to indium content. Fig.\ \ref{tem_fig}c shows a large scale \gls{haadf}-\gls{stem} image of P24 with contrast fluctuations across the entire sample, seemingly having no correlation with any individual part of the structure. These fluctuations are almost certainly an artefact: comparing indium content extracted from \gls{eds} spectroscopy with the \gls{haadf} profile across the \gls{qw} (Fig.\ \ref{tem_fig}d) near a typical fluctuation, we observe that the \gls{haadf} contrast fluctuation does not correspond to any change in indium concentration. These fluctuations may be due to surface relaxation or contamination from sample preparation; at an acceleration voltage of only 80 kV, the electron beam is sensitive to any potential surface effects. To avoid this artefact, we rely on \gls{eds} measurements rather than \gls{haadf} images: Figs.\ \ref{tem_fig}e \& f display indium content profiles across the \glspl{qw} of P0 and P24, obtained by averaging the \gls{eds} signal over the regions in Figs.\ \ref{tem_fig}a \& b. Indeed there is a high plateau in the profiles exactly three \glspl{ml} thick, but the \gls{qw} top interface is not sharp---a characteristic of indium segregation during growth.

To quantify the indium surface segregation, we can apply a model developed by Muraki \textit{et al.}\cite{Muraki1992} In this model, during the growth of one molecular \gls{ml} of InGaN a certain fraction $(1-R)$ of the indium atoms at the surface are incorporated into the \gls{ml} while the remaining fraction $R$ segregates to the new surface. As such, $R$ represents the degree of segregation that occurred during growth. The In content in the \textit{n}\textsuperscript{th} \gls{ml}, $x_n$, is then given by:

\begin{align}
	x_n &=x_0 (1-R^n) &(1\leq n\leq N_{\textnormal{QW}}; \textnormal{well}),\\
	x_n &=x_0 (1-R^{N_{\textnormal{QW}}})R^{n-N_{\textnormal{QW}}} &(n> N_{\textnormal{QW}}; \textnormal{barrier}),
\end{align}

\noindent where $N_{\textnormal{QW}}$ and $x_0$ are the nominal width and indium content of the \gls{qw}, respectively. This model was fitted to the \gls{stem}-\gls{eds} results for the \glspl{qw} of P0 and P24, as shown in Figs.\ \ref{tem_fig}e \& f. Only the data for which the measured In content was greater than 5 \% were considered for the fitting process; this was to avoid the $\sim 1$ \% zero error present in the \gls{eds} measurement. The zero error arises from noise in the \gls{eds} spectra which is falsely quantified when the indium content is below the detection limit. The derived fit could then be extrapolated into the spurious regions, eliminating the error in subsequent analyses. $N_{\textnormal{QW}}$ was the same for both samples at 4.3, close to the target value of 3; meanwhile $x_0$ was 26 \% for both samples. The values of $R$ extracted (0.82 and 0.86) are in line with previous estimations of the segregation coefficient in InGaN/GaN \glspl{qw}.\cite{Dussaigne2003} Crucially, the \gls{qw} of P0 possesses a slightly larger $R$ value than that of P24, indicating that greater indium segregation occurred in this sample. This could explain the $\sim66$ \si{\milli\electronvolt} \glsentryshort{pl} redshift between P0 and P24 seen in Fig.\ \ref{sample_info}c of the main text---to confirm, we carried out Schrödinger-Poisson calculations.

\begin{table}[h!]
	\caption{ \normalsize \label{simresults}Comparison of P0 and P24 emission energies gained from room-temperature \glsentryshort{pl} measurements (experiment) versus transition energies predicted by \textit{nextnano} Schrödinger-Poisson calculations (simulation) at 300 K. The difference between the experimental/simulation energies of P0 and P24 are given in the final column.}
	\begin{ruledtabular}
		\begin{tabular}{lccr}
			&P0 energy (eV) & P24 energy (eV) & Difference $\Delta E_{\textnormal{P0-P24}}$ (meV)\\
			Experiment & 3.006 & 2.940 & 66\\
			Simulation & 3.117 & 3.068 & 49
		\end{tabular}
	\end{ruledtabular}
\end{table}

For these calculations, the entire structures of P0 and P24 were simulated at 300 \si{\kelvin} using \textit{nextnano},\cite{Birner2007} including the accurate \gls{ml}-by-\gls{ml} \gls{qw} indium profiles obtained from the Muraki fit. We can compare the simulated results for P0 and P24 to those gained from \glsentryshort{pl} at room temperature using a HeCd laser (Table \ref{simresults}). The difference in transition energies between the \glspl{qw} of P0 and P24 is predicted to be 49 \si{\milli\electronvolt} at room temperature, which accounts quite well for the $\sim66$ \si{\milli\electronvolt} \glsentryshort{pl} redshift observed between P0 and P24, supporting our structural analysis. However, a noticeable discrepancy exists between the experiment and simulated absolute values for each sample ($\sim110$ \si{\milli\electronvolt}). Multiple factors contribute to this discrepancy. Firstly, the simulation does not consider the exciton binding energy in the \gls{qw}, which lowers the emission energy relative to the calculated transition energy. Taking GaN as an example, in the limiting exact 2D case of an infinitely deep \gls{qw} the exciton binding energy is increased by a factor of four relative to the bulk value, placing it at around 100 \si{\milli\electronvolt}.\cite{Vurgaftman2003} Secondly, a Stokes shift is expected between absorption and emission energies due to tail states within the bandgap; this effect is further exacerbated by alloy disorder in the InGaN/GaN \gls{qw}.\cite{Glauser2014} Finally, \gls{eds} quantification is never absolutely precise, and there is at least a $\pm 1$ at.\% systematic error involved. All of these factors should be roughly constant for both samples, and hence do not impede the comparison between them.  

\begin{figure}[h]
	\centering
	\includegraphics[width=1\columnwidth]{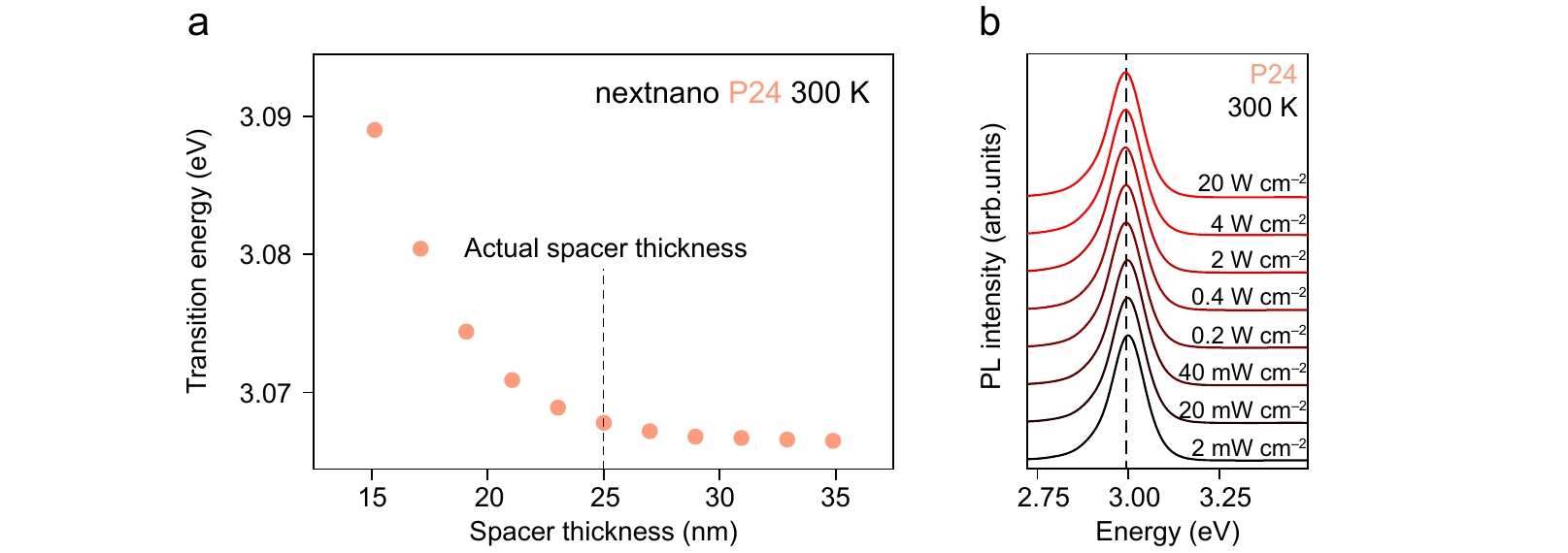}
	\caption{ \normalsize \textbf{a} Transition energies calculated by \textit{nextnano} for the P24 \gls{qw} as the GaN spacer thickness between the SL and the QW was varied. The temperature was set to 300 K. \textbf{b} \glsentryshort{qw} \glsentryshort{pl} series on P24 at 300 \si{\kelvin} obtained using a 325 \si{\nano\metre} HeCd laser, varying the power density from 2 \si{\milli\watt\per\centi\meter\squared} to 20 \si{\watt\per\centi\meter\squared}. Spectra are individually normalised and offset along the \textit{y}-axis for clarity. No shift in peak energy occurs across the entire power density range.}
	\refstepcounter{SIfig}\label{power}
\end{figure}

Finally, it was worth ruling out any effect on the fundamental \gls{qw} transition energy from the electric field induced by the InAlN/GaN \gls{sl}; as can be seen in Fig.\ \ref{power}a, at our GaN spacer thickness of 25 \si{\nano\metre} between the \gls{sl} and \gls{qw} the electric field of the \gls{sl} has been mostly screened. Increasing the spacer thickness to 35 nm would result in a change in the transition energy of only $\sim1$ \si{\milli\electronvolt}. Furthermore, the absence of any impact from the built-in electric field can be verified experimentally by measuring \glsentryshort{pl} spectra for P24 under various excitation power densities (Fig.\ \ref{power}b). Any effect from the electric field would be revealed as a blueshift in \gls{qw} emission energy at higher power densities due to carriers screening the field. Conversely, we see that the peak emission energy remains constant across the entire power range, confirming that the 66 \si{\milli\electronvolt} discrepancy in peak emission energy between P0 and P24 is not due to any electric field from the \gls{sl}. Furthermore, this clearly demonstrates the limited effect of the quantum-confined Stark effect in our ultra-thin \glspl{qw}; consequently, we do not need to consider any field-screening effects in in our probe current study (Fig.\ \ref{current} of the main text).

In summary, this complete analysis ensures we can be confident in the segregated indium profiles measured by \gls{eds} and their corresponding Muraki fits. The analysis identifies marginally greater segregation in the \gls{qw} of P0 compared to P24, which explains the observed \gls{pl} peak energy difference between the samples (with any effect from the built-in electric field being ruled out experimentally and theoretically). Slight changes in segregation can then explain the small variances in peak energy between the samples seen in the room-temperature \gls{qw} \gls{pl} (Fig.\ \ref{sample_info}c of the main text). For our calculation of \gls{pd} 3-D densities (\si{\per\centi\metre\cubed}) from 2-D values (\si{\per\centi\metre\squared}), we use the average \gls{fwhm} of the \gls{eds} indium profiles at $\sim1.8$ nm.

\section{COMPLETE DIFFUSION ANALYSIS}
\label{comp_diff_sec}

\begin{figure}[h]
	\centering
	\includegraphics[width=1\columnwidth]{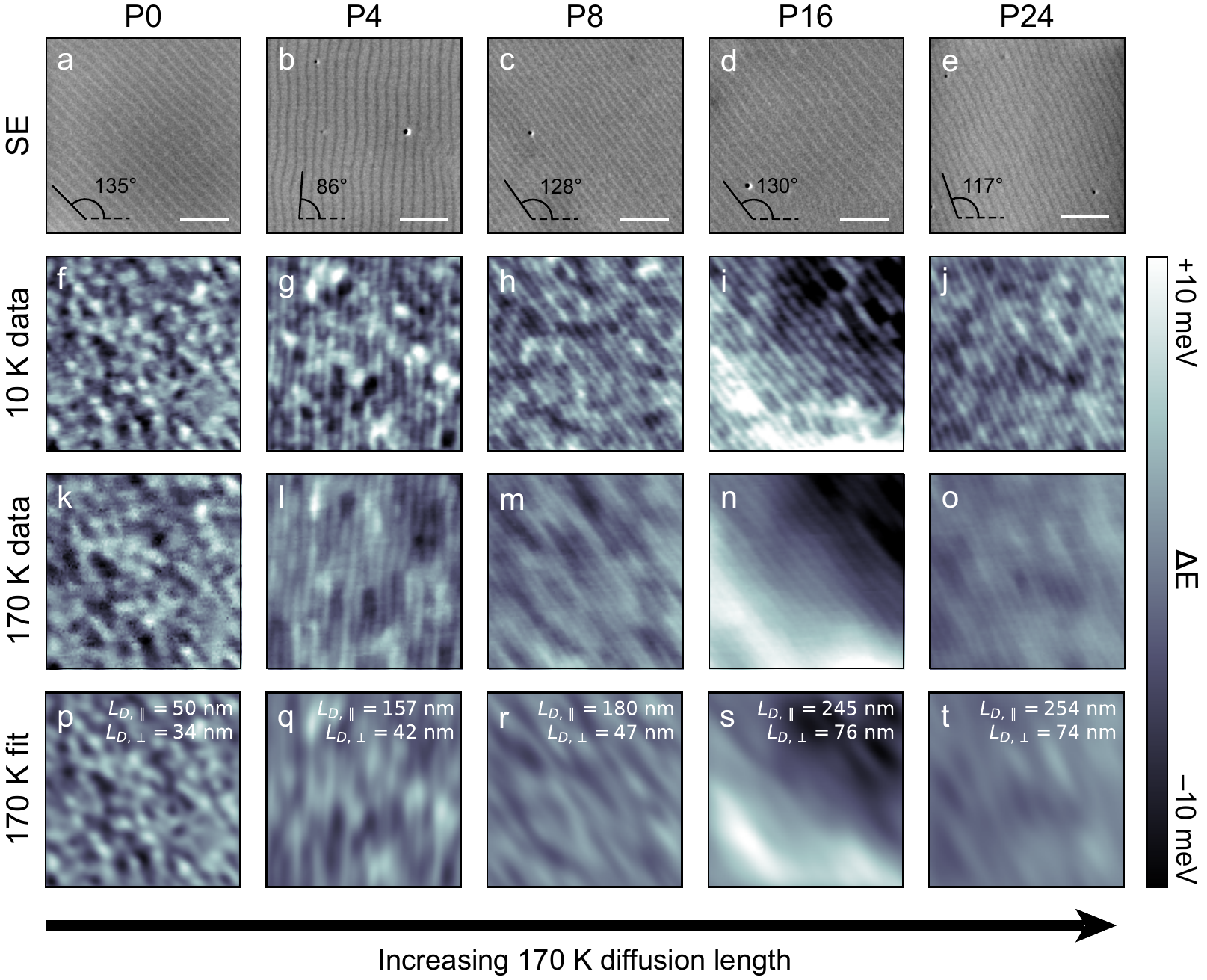}
	\caption{ \normalsize \textbf{(a--e)} \glsentryshort{se} images for all samples as presented in the main text (Figs.\ \ref{pd_cl}a--e) with the step-edge direction marked by an angle as calculated by Fourier transform analysis (scale bars 500 nm). \textbf{(f--j)} 10 K and \textbf{(k--o)} 170 K \glsentryshort{qw} \glsentryshort{cl} peak energy images from the same areas as in the \glsentryshort{se} images. Peak energy is plotted relative to the mean energy of each image as in the main text.  \textbf{(p--t)} Images obtained from the Gaussian convolution procedure applied to the 10 K \glsentryshort{cl} images in f--j, with calculated diffusion lengths indicated.}
	\refstepcounter{SIfig}\label{diffusion}
\end{figure}

In Fig.\ \ref{diff}e of the main text we presented the diffusion lengths calculated for all samples from the Gaussian convolution fitting method. Here we show the peak energy images used to obtain these diffusion lengths in Fig.\ \ref{diffusion}. Step-edge directions were calculated by Fourier transform analysis of the \gls{se} images (Figs.\ \ref{diffusion}a--e). The impact of the step-edges on the peak energy is evident for P4--P24 in Figs.\ \ref{diffusion}g--j, which leads to strongly anisotropic diffusion in all of these samples. This is confirmed by the diffusion lengths parallel ($L_{D, \parallel}$) and perpendicular ($L_{D, \perp}$) to the step-edges extracted by Gaussian convolution (Figs.\ \ref{diffusion}p--t). These convoluted images are in good agreement with the real 170 K peak energy images (Figs.\ \ref{diffusion}k--o), confirming the validity of our simple diffusion analysis. The error bars in diffusion length shown in Fig.\ \ref{diff} of the main text were estimated by varying the initial parameters of the Gaussian convolution fit.

\clearpage

\twocolumngrid
\pdfbookmark[0]{References}{sec:ref}

\textbf{References}

\normalsize

\bibliography{References}

\end{document}